%% file: SHOCK_7a.tex
\documentclass[showpacs,showkeys,preprintnumbers,amsmath,amssymb,times,graphicx]{revtex4}

\usepackage{graphicx} 
\usepackage{dcolumn}
\usepackage{bm}

\usepackage{color}
\usepackage{amsmath}
\usepackage{amsfonts}
\usepackage{amssymb}
\usepackage{amsbsy}
\usepackage[figuresright]{rotating}
\usepackage[font=small,labelfont=bf]{caption}
\usepackage{natbib}

\input{calbook_a_defs}

\begin{document}

\title{Planetary Bow Shocks}

\author{R. A. Treumann$^\dag$ and C. H. Jaroschek$^{\ddag}$}\email{treumann@issibern.ch}
\affiliation{$^\dag$ Department of Geophysics and Environmental Sciences, Munich University, D-80333 Munich, Germany  \\ 
Department of Physics and Astronomy, Dartmouth College, Hanover, 03755 NH, USA \\ 
$^{\ddag}$Department Earth \& Planetary Science, University of Tokyo, Tokyo, Japan
}%

\begin{abstract} Our present knowledge of the properties of the various planetary bow shocks is briefly reviewed. We do not follow the astronomical ordering of the planets. We rather distinguish between magnetised and unmagnetised planets which groups Mercury and Earth with the outer giant planets of the solar system, Mars and Moon in a separate group lacking magnetic fields and dense atmospheres, and Venus together with the comets as the atmospheric celestial objects exposed to the solar wind. Asteroids would, in this classification, fall into the group together with the Moon and should behave similarly though being much smaller. Extrasolar planets are not considered as we have only remote information about their behaviour. The presentation is brief in the sense that our in situ knowledge is rather sporadic yet, depending on just a countable number of bow shock crossings from which just some basic conclusions can be drawn about size, stationarity, shape and nature of the respective shock. The only bow shock of which we have sufficient information to deal in sufficient depth with its physics is Earth's bow shock. This has been reviewed in other places in this volume and therefore is mentioned here only as the bow shock paradigm in passing.\end{abstract}
\maketitle 

\section{Introduction}
\noindent Among the shocks in the heliosphere planetary bow shocks occupy an exclusive position due to their comparably easy accessibility. It is in particular the Earth's bow shock that has provided the deepest insight into the structure and formation of collisionless shocks. Much of the physics that was learnt from its continuous observation by spacecraft crossing it multiply has been dealt with in the chapters of Part I of this volume. In the present chapter we will provide a brief review of some of the new developments in planetary bow shock physics that have not been mentioned yet and go beyond the existing excellent review paper of \cite{Russell1985} and the short and more specialised papers contained in Section 6 on Planetary Shocks in \cite{Russell1995}.

When speaking about planetary bow shocks we have in mind all the different kinds of shocks which are generated when a heavenly solid obstacle is put into the high speed flow of the solar wind. Since the dominant such obstacles are the planets, planetary bow shocks also dominate this selection even though they are not just the most abundant in the solar system and heliosphere.  The more abundant ones are the bow shocks around the many solid asteroids which occupy some parts of interplanetary space sometimes passing the vicinity of Earth in the inner region of the solar system, each of them creating its own bow shock. 

These asteroidic bow shocks are, however, of small sizes and thus do not attract our main attention. Since the asteroids do not have atmospheres and at least to our knowledge no magnetic fields either, their bow shocks are of similar nature as the bow shocks of other non-magnetised solid bodies in the solar system like most planetary satellites, in particular like the moon, and can thus considered to be smaller relatives of the lunar bow shock. There is, however, one distinction which is indeed related to the very size of these objects. The physics of their bow shocks changes completely relative to those like the lunar bow shock if the size of the asteroid becomes comparable to the ion inertial length. In this case -- possibly -- no bow shock can evolve at all as from the plasma point of view the obstacle (asteroid) becomes a very heavy, possibly charged and (due to the solar wind induction effect) magnetised very large ``particle". 

Other heavenly bodies which pass the solar system are comets. These are non-magnetic but gaseous objects. As a consequence the bow shocks developing around them differ from those of the magnetised planets but have much in common with non-magnetic planets possessing dense atmospheres like Venus with the main difference of the lacking gravity, the completely different composition of the atmosphere, and the size as comets are bodies that are as small as asteroids. Nevertheless, from the plasma point of view they cannot be considered to behave like particles just because of the outgassing of their atmosphere which forms an atmospheric gaseous cloud around the cometary body that is several orders of magnitude larger in extent than the cometary body.

In the following we will therefore consider three classes of planetary bow shocks: bow shocks around magnetised planets, cometary bow shocks including the Venusian bow shock, and the lunar bow shock (as a paradigm for other non-magnetised non-atmospheric  bow shocks). 

We are interested in the physics of shocks here. In the present case this means that we are interested in the physics of bow shocks under the various differing external plasma and planetary atmospheric conditions.  While the available information about the known planetary bow shocks decreases steeply with distance from Earth, the presentation will naturally be rather short, sticking just to the few conclusions that can be drawn from the sporadic measurements that are available to us from the few spacecraft crossings {\it in situ} of the more remote planetary bow shocks. 
\begin{figure}[t!]
\centerline{\includegraphics[width=0.80\textwidth,clip=]{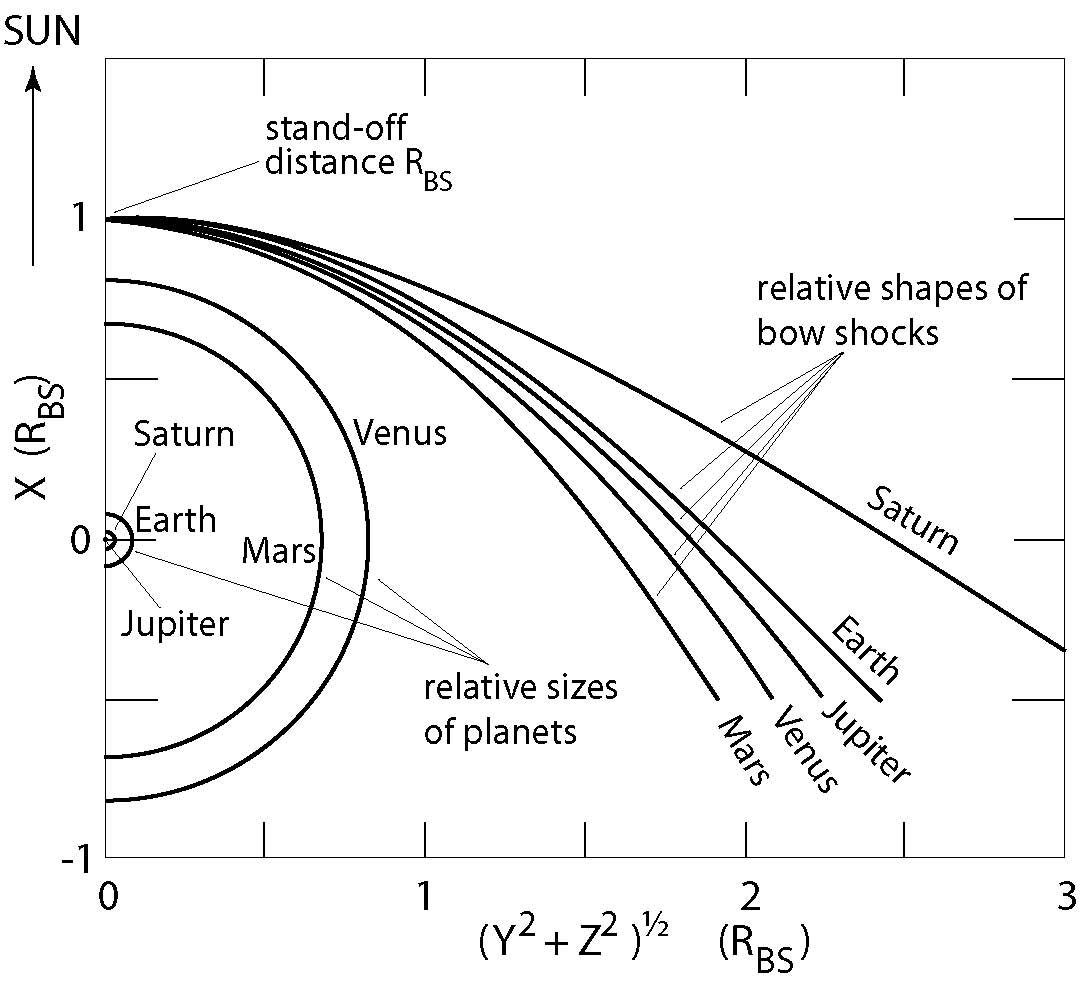}}
\caption[] 
{\footnotesize The semi-theoretical shapes of some planetary bow shocks. Length scales have been normalised to the stand-off nose distances $R_{\rm BS}$ of the corresponding planetary bow shocks. In this representation the sizes of the planetary bodies are inverted. Jupiter shrinks to a point at the origin while the small planets Mars and Venus become grossly oversized.  Mercury is not yet included here. Of the outer giant planets only Saturn is shown. \citep[data taken from][]{Slavin1985}.}\label{chapBS-fig-Slavin}
\end{figure}

Earth's bow shock is the only bow shock that has been sufficiently monitored to conclude about the formation and internal physics of a bow shock. This has been made use of in various different places in this volume and in various other publications. Therefore, we are treating Earth's bow shock here just as the paradigm of a magnetised planetary bow shock, and this we will do only briefly. We are less interested in the shapes and gasdynamic properties of bow shocks as they do not tell us very much about the structure, internal dynamics and dissipation processes acting inside a bow shock. Shapes and gasdynamic properties of bow shocks have been reviewed in depth in other papers \citep[see, for instance,][among others]{Slavin1985,Russell1985,Russell1995,Spreiter1995}, based on similarities and scalings to Earth's bow shock and the (relatively meagre) statistics available from the small number of planetary bow shock crossings. Figure \ref{chapBS-fig-Slavin} provides an overview of some of the planetary bow shock shapes properly scaled to their stand-off distances from the centres of their respective planets as obtained from the assumption of similarity to Earth's bow shock and  the available knowledge from the locations of the sporadic spacecraft bow shock crossings.

\section{Terrestrial Type Bow Shocks}\index{shocks!terrestrial type bow shocks}
\noindent The terrestrial planets are Earth, Mercury, Venus, Mars, and -- though not a planet -- a body of similar consistence: the Moon. This classification applies to their interior structure, composition and densities. However, from the bow shock viewpoint the classification does not follow this scheme. Here one rather distinguishes the magnetised and non-magnetic planets which leads to group together Earth with Mercury and the giant magnetised planets, Jupiter, Saturn, Uranus, and Neptune. Surprisingly  one of the Jovian moons, Ganymede,  has an own magnetic field and could also be included. However, like the other satellites of the giant planets including Titan, Ganymede\index{moons!Ganymede} never leaves Jupiter's magnetosphere \index{magnetosphere!Jupiter} to enter the solar wind. It possesses a magnetosphere which in the Jovian magnetospheric plasma flow stands up as an obstacle. Therefore a bow shock he would possess would be of purely Jovian origin and required super-Alfv\'enic flow in the Jovian magnetospheric plasma. Since this could only occasionally be the case we do not include Ganymede or any other close planetary satellite (with the exception of the Moon) into this collection.

\subsection{Earth's Bow Shock}\label{chapPBS-sec-EBS}
\noindent The paradigm of a planetary bow shock embedded into a super-magnetosonic stellar wind flow is the Earth's bow shock wave. Its continuous observation has fertilised the understanding of collisionless shock physics like no other shock neither in space or in the laboratory. The fortunate properties of the bow shock -- we are simply speaking of the bow shock if speaking of Earth's bow shock -- are multitude: it is a supercritical shock, it is a magnetised shock, it is a curved shock, and because of this latter property it at one and the same time possesses regions of perpendicular, quasi-perpendicular, oblique, and quasi-parallel nature permitting to study all the properties that these different shocks develop separately and in their relation. In addition it is a sufficiently extended shock of tangential diameter $\gg \Delta_s, \lambda_i, r_{ci}$ that by far exceeds its width $\Delta_s$, the ion inertial length $\lambda_i$ and the ion gyro-radius $r_{ci}$. 

Macroscopically the bow shock reacts sensitively to the variations in the upstream flow, the upstream turbulence, direction of the interplanetary magnetic field and to macroscopic disturbances in the upstream flow thereby permitting to study these dependencies and variations and the different states of a collisionless shock that are caused by the variations in the upstream plasma parameters like the respective speeds, temperatures and densities of the flow, the composition of the upstream plasma, and the upstream state of turbulence. It also allows to study the downstream properties of the shocked plasma behind the different quasi-perpendicular and quasi-parallel parts of the shock, the development and properties of the downstream turbulence, particle dynamics and the downstream mixing state. This is due to the blunt nature of the obstacle that is responsible for the existence of the bow shock, Earth's magnetosphere \index{magnetosphere!Earth} with its comparably strong geomagnetic surface magnetic field of $B_{\rm E}(1\,{\rm R_E})\sim 4.3 \times 10^{4}$ nT that is much larger than the average solar wind magnetic field $B_{\rm SW}(1\,{\rm AU})\sim 10$ nT and therefore exerts a strong resistance to the solar wind flow, digging a large cavity into the flow and forcing it to deviate from its original direction to turn around this cavity, the magnetosphere. This bluntness of the magnetosphere gives sufficient space in upstream direction between the magnetosphere and  bow shock for the accumulation of shock disturbed plasma. 

The bow shock that is created in this interaction between the solar wind flow and the geomagnetic field is located at an upstream distance that follows from hydrodynamic theory as the envelope of the characteristics of waves that can propagate upstream against the flow in the time the flow needs to pass from the shock to the magnetosphere. For the bow shock this distance at the nose of the bow shock in solar direction is in the average  found at a geocentric radius of $R_{\rm BS}\sim (13-14)\,{\rm R_E}$. It is of approximately hyperbolic shape widening with distance from the Earth-Sun line in such a way that its radius at the terminator becomes approximately $\sim 27\,{\rm R_E}$. 

A simple formula that approximately describes the bow shock nose location distance $\Delta_d=R_{\rm BS}-R_{\rm MP}$ from the nose location of the magnetosphere $R_{\rm MP}$, i.e. the magnetopause distance along the Earth-Sun line at local noon, as function of the shock density compression ratio $N_2/N_1$ has early  been given  by \cite{Spreiter1966} from their gasdynamic calculation as 
\begin{equation}\label{chapPBS-eq-BSdist}
\Delta_d\approx 1.1 \frac{N_1}{N_2}R_{\rm MP},\qquad \frac{N_1}{N_2}\approx \frac{\gamma+1}{\gamma-1}-2\frac{\gamma+1}{{\cal M}^2(\gamma-1)^2}
\end{equation}
where $\gamma$ is the adiabatic index, and ${\cal M}$ the (magnetosonic) upstream Mach number, which in the hydrodynamic approximation used is the ordinary gasdynamic Mach number. The upstream magnetic field is not included here since the magnetic pressure in the upstream solar wind is small and in this early theory not much was known about the quasi-parallel and quasi-perpendicular differences between shocks. However, what concerns the nose distance the magnetic effect on the global form is not so important. It is usually incorporated into a constant factor $K\sim O(1)$ in the definition of the magnetopause nose distance $R_{\rm MP}$ which is defined in terms of the planetary radius $R_{\rm P}$ for a planetary dipole magnetic field of surface strength $B_{\rm P}$ through the magnetic and dynamic pressure balance as
\begin{equation}
R_{\rm MP}=\left(\frac{KB_{\rm P}^2}{m_i\mu_0N_iV_1^2}\right)^{\!\!\frac{1}{6}} R_{\rm P}
\end{equation}
The factor in parentheses is the square of the Alfv\'en speed based on the planetary surface magnetic field, the upstream solar wind density $N_i$ and the upstream velocity $V_1$. The sixth root makes this distance only weakly dependent on the factor $K$. The relative variation of the distance $\Delta_d$ between the magnetopause and bow shock is given by 
\begin{equation}\label{chapPBS-equ-Deltad}
\frac{1}{\Delta_d}\frac{{\rm d}\Delta_d}{{\rm d}{\textsf P}_{\rm SW}}=-\frac{1}{6{\textsf P}_{\rm SW}}
\end{equation}
where ${\textsf P}_{\rm SW}=m_iN_iV_1^2$ is the ram pressure of the solar wind.

The above formulae suggest the intuitively reasonable result that an increase in the compression ratio will push the bow shock closer to Earth. For the largest MHD compression ratio of $N_2/N_1=4$ this distance is $\sim 0.275\,R_{\rm MP}$. At the Earth with $R_{\rm MP}\approx 10\,{\rm R_E}$ the bow shock nose is at a geocentric distance of $R_{\rm BS}\approx 12.75\,{\rm R_E}$. Because of the assumption of maximum compression ratio this is somewhat below the average distance noted above. But the above handy formula is not solely restricted to Earth's bow shock; it can be applied to any other planetary bow shock as long as the magnetic effects on the shape and dynamics of the shock can be neglected which is the case as long as $\beta_u\ll 1$, where $\beta_u= \mu_0 m_iN_1V_1^2/B_1^2$ is the non-relativistic upstream plasma $\beta$ that is based on the kinetic energy of the flow. Figure \ref{chapBS-fig-Slavin} shows the semi-theoretical relative shapes of the bow shocks for some planets where all lengths have been normalised to the theoretical stand-off distances of the corresponding planetary bow shocks. In this representation the sizes of the various planets are inverted. Jupiter shrinks to a point at the origin while Mars and Venus become oversized.  

These expressions can be applied to any planet immersed into the solar wind stream under the condition that its magnetic field is to good approximation a dipole field. The solar wind velocity $V_1$ is a constant throughout the solar planetary system. However, the solar wind density changes about like $N_i\propto R^{-2}$ such that the solar wind dynamical pressure decreases with heliocentric distance. In application to other planets it is therefore convenient to measure $R$ in unit of 1 AU and to use the average solar wind density at Earth's orbit as reference. This density is in the average $N_i=5\times 10^6\,{\rm m}^{-3}$.
\begin{figure}[t!]
\centerline{\includegraphics[width=1.0\textwidth,clip=]{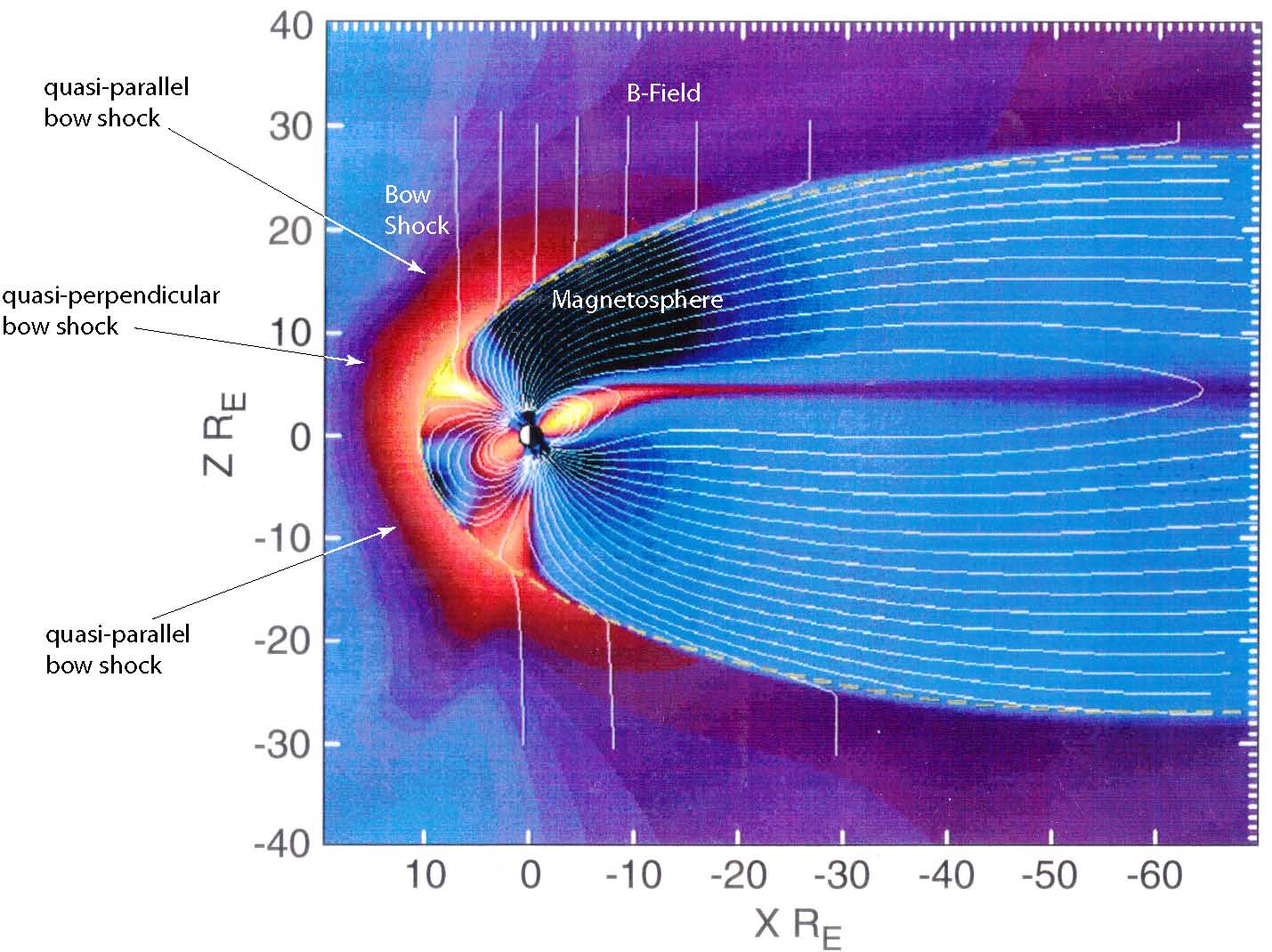}}
\caption[] 
{\footnotesize View of the whole magnetosphere and bow shock system embedded in the solar wind stream. The figure is obtained by application of the empirical magnetospheric Tsyganenko-model that is based on observations. Shown is the cross section that includes the Sun-Earth line, terminator and rotation axis of Earth for summer on the northern hemisphere and for a Nort-South directed magnetic field vertical to the excliptic  being completely reconnected to the magnetospheric magnetic field. The sun is on the far left. The bow shock is the sunward edge of the red region on the left in front of the magnetosphere. Colouring indicates  plasma temperature/thermal pressure with red highest and dark lowest.}\label{chapBS-fig-msph}
\end{figure}

Figure\,\ref{chapBS-fig-msph} gives an impression of the bow shock-magnetosphere system as obtained from the empirical data based Tsyganenko model for the particular case of northern hemispheric summer and strictly north-south directed solar wind magnetic field such that the field is completely reconnected to the geomagnetic field inside the magnetosphere. The solar wind is streaming from the left against the geomagnetic field creating the magnetosphere and causing the bow shock in front of the magnetosphere. 

The figure shows the asymmetry of the bow shock and magnetosphere that is caused by the inclination of Earth's rotation axis against the ecliptic. Slowing down of the solar wind flow and downtail reacceleration of the plasma is recognisable from the relative distance between adjacent magnetic field lines. Behind the bow shock the distance between the field lines is small. When the downstream plasma couples newly to the solar wind the distances between the field lines increase along the magnetospheric tail in anti-solar direction. Colouring indicates the plasma temperature and thermal pressure. It shows the transition from the cool solar wind flow across the shock to the hot and dense plasma behind the shock in the magnetosheath, with the bow shock being the boundary between both plasmas, itself being responsible for the heating and compression of the plasma and magnetic field.

The figure also shows that even in this case when the magnetic field is exactly perpendicular to the ecliptic plane the bow shock is bent in a way that two large regions of quasi-parallel shocks surround the nose region which is quasi-perpendicular. This property of the bow shock and its practically continuous and uninterrupted presence make it so attractive for the investigation of the properties of supercritical collisionless quasi-perpendicular as well as quasi-parallel shocks. This is why most of our knowledge of the structure and physics of collisionless supercritical shocks has been stimulated by investigations of the bow shock. In the following we briefly describe the global conditions prevalent at the terrestrial bow shock before turning to the discussion of the differences and similarities between the Earth's bow shock and the bow shocks encountered near the other heavenly bodies in our solar system.
\begin{table}[t]
\caption{$\quad$Solar Wind Conditions Upstream of the Planets}\index{solar wind!upstream conditions}
\begin{center}
\begin{tabular}{l|rr|rrrcc}
\hline
& & & & &&&\\[-0.3ex]
Planet& $R{~~~}$ & Spin $T_{\rm P}^\dag$ & $10^{-9}{\textsf P}_{\rm SW}$ & ${\cal M}_s^*$ & ${\cal M}_A^*$ & $\beta$ & ${\bf B}_{\rm SW}$-angle \\
            & AU{~}   & hrs{~~~~} & ${\rm J\,m}^{-3}{~~}$ &                    &                     &                   &   $\alpha^\circ_{\rm Parker}$  \\ [1.5ex]           
\hline
& & & & & & &\\
Mercury&0.31&1407.50&26.5000&5.5&3.9&0.5&17 \\[0.3ex]
            & 0.47&& 11.0000 &6.1 &5.7&0.9&25 \\[-0.3ex]
Venus&0.72&-5832.50 & 5.0000&6.6&7.9&1.4&36\\
Earth &1.00&23.93& 2.5000&7.2&9.4&1.7&45\\ 
Moon & & 655.73& & & & &  \\
Mars&1.52&24.62 &1.1000&7.9&11.1&2.0&57\\
Jupiter&5.20&9.93&0.0920&10.2&13.0&1.6&79\\
Saturn&9.60&10.49&0.0270&11.6&13.3&1.3&84\\
Uranus&19.10&-17.24&0.0069&13.3&13.3&1.0&87\\
Neptune&30.20&16.11&0.0027&14.6&13.3&0.8&88\\ [1ex]
 \hline 
 \multicolumn{8}{l}{} \\[-0.7ex]
\multicolumn{8}{l}{$^*${\footnotesize Respective sonic (subscript $s$) and Alfv\'enic (subscript $A$) Mach numbers.}} \\[-0.3ex]
 \multicolumn{8}{l}{$^\dag${\footnotesize Spin period of planet in terrestrial hours. The sign indicates direction of rotation.}} \\[-0.3ex]
\multicolumn{8}{l}{{\footnotesize ~~Solar wind data taken from \cite{Slavin1981}.}}
 \end{tabular}
\end{center}
\vspace{-0.5cm} 
\label{chapPBS-table1}
\end{table}
\subsubsection{Solar wind conditions}
\noindent Compared to Earth's bow shock  the information obtained on the remaining planetary bow shocks is still sparse. In order to be able to classify the global properties of the true planetary bow shocks, i.e. the bow shocks of the planets and their satellites, one needs to know the respective upstream solar wind conditions. In other words one needs to know the variation of the main properties of the solar wind with heliocentric distance. These are given in Table \ref{chapPBS-table1} for the radial heliocentric distance $R$  of the planets measured in AU. Mercury, the planet with orbit closest to the sun, has two entries because of its large orbital eccentricity. The decrease in solar wind ram pressure ${\textsf P}_{\rm SW}$ with heliocentric distance is due to the $N\sim R^{-2}$ solar wind dependence on radius. ${\cal M}_s$ and ${\cal M}_A$ are the respective sonic and Alfv\'enic Mach numbers, $\beta=2\mu_0NT/B^2$ is the ratio of thermal to magnetic pressures. Since $B\sim R^{-1}$, the radial dependence of $\beta$ is mainly determined by the radial variation of the solar wind temperature $T=T_i+T_e\approx T_e$; outside the Mars orbit the solar wind cools about adiabatically. Finally, the spiral angle of the interplanetary magnetic field ${\bf B}$ against the radial direction of the solar wind flow in the ecliptic plane increases gradually to become about perpendicular to the solar wind already at the orbits of the outer giant planets. This table is built on both theory and observation by the {\small Helios} and {\small Mariner 10} spacecraft.\index{spacecraft!Helios}\index{spacecraft!Mariner 10}

In addition the solar wind exhibits a susceptible level of turbulence, fluctuations in the magnetic field, density and flow velocity, dependence on the solar cycle and the well known magnetic sector structure. Occasionally, when the planet passes a sector boundary the magnetic field direction may change abruptly and the planetary bow shock may be subjected to the passage of a current sheet which affects the structure of the planetary bow shock and creates effects like the formation of High Flow Anomalies (HFA). Moreover, Coronal Mass Ejections (CME) cause interplanetary shocks to pass over the planet and interact with the bow shock. These effects are transient, however, temporarily changing the character, type, shape and structure of the bow shock.

\subsubsection{Properties}\index{shocks!bow shock properties}
\noindent The Earth's bow shock has served as the paradigm of a supercritical shocks in the previous chapters of this volume. We may briefly summarise its various properties as follows:

The Earth's bow shock is a permanent feature of the solar wind-magnetosphere system, standing in front of the magnetosphere at an average geocentric nose distance of $R_{\rm BS}=(12-14)\,{\rm R_E}$, a distance that may change substantially depending on the variations of the solar wind pressure, i.e. depending on the compressional state of the Earth's magnetosphere. Occasionally the bow shock can be found at a distance as close as $\lesssim 10$\,R$_{\rm E}$ or also much farther out at nose distances $\gtrsim 15$\,R$_{\rm E}$. Its shape is about hyperbolic forming a shield in front of the magnetosphere that opens up to the flanks and poles.  Since its shape is convex around the magnetosphere when looked at it from the sun, it has at least one spot on its surface where the solar wind magnetic field is tangential and the shock is perpendicular, respectively quasi-perpendicular.  The location of this perpendicular spot and the quasi-perpendicular part around it depends on the spiral angle of the magnetic field. Large parts of the bow shock towards its flanks are quasi-parallel.

The bow shock is highly supercritical with Mach number $6\lesssim {\cal M}\lesssim 12$, occasionally even larger. It is a strong shock in the sense that the shock compression ratio is close to what is know to be the maximum of the compression ratio of a supercritical MHD shock $2.5< N_2/N_1\sim B_2/B_1<4$. This ratio is based on the values of the fields well in front and well behind the shock. When using the overshoot fields the compression ratio can be higher than the MHD maximum. This is attributed to kinetic effects that take place in the shock ramp and transition region on scales $<\lambda_i$, shorter than the ion inertial length where electron kinetics comes into play. The width of the shock transition is of the order of $1\lesssim \Delta_s/\lambda_i\lesssim (3-4)$, depending on whether one is talking of the quasi-perpendicular or quasi-parallel bow shock. The width of the quasi-perpendicular shock ramp is usually at the lower end of this interval, $\Delta_{\rm ramp}\sim \lambda_i$ or less. 

The quasi-perpendicular bow shock reflects a substantial amount of ions back upstream into the solar wind and already retards the solar wind by a fraction of $\sim (10-30)\%$ in velocity. It possesses a foot region which has been measured to be slightly broader than one ion gyroradius based on the upstream field, $\Delta_{\rm foot}\sim r_{ci}$. This foot is caused by the current carried by the reflected particles in the foot which are subjected to acceleration in the upstream motional electric field that is locally tangential to the shock surface. Observationally it has not yet been identified to what extent  the reflected, heated and accelerated electrons contribute to this current. It is also not precisely known which instability is responsible for the shock reformation. Full particle simulations with realistic mass ratio suggest -- without experimental proof --  that the instability is dominated by the electrons, being of the type of modified two-stream instability. Reformation might then be driven by electrons which for high Mach numbers can become reflected from localised electric fields in combination with the magnetic mirror force. 
\begin{table}[t]
\caption{$\quad$Relevant Planetary Properties}\index{planets!planetary properties}
\begin{center}
\begin{tabular}{l|rrccrrr}
\hline
& & & &&&&\\[-0.3ex]
Planet & $R_{\rm P}{~}$ &  $M_{\rm P}{~~~~}$ & $M_{\rm P}/M_{\rm E}$ & $|{\bf B}({\rm R_P})|^\ddag$ & ${\bf B}$-Tilt & $R_{\rm MP}$ & $R_{\rm BS}$\\
            & km    &  T\,m$^3${~~}  & & $\mu$T   &  $\theta^\circ{~~}$  &    R$_{\rm P}{~}$ & R$_{\rm P}{~}$\\ [1.5ex]           
\hline
& &  & & & &&\\
Mercury&2440  &$6\cdot10^{12}$& $4\cdot10^{-4}$& $4.15$& &$<$1.2$^\dag$& $\sim$1.5\\[0.3ex]
            &  &&&&&1.37$^\dag$& \\[-0.3ex]
Venus&6051&$1.26\cdot10^{11}$&$1.3\cdot10^{-4}$&$0.001$&&0.56$^*$&\\
Earth &6378&$9.9\cdot10^{16}$&1&31&+11.3&$\sim$11&\\ 
Moon & 1738& $1.6\cdot10^{9}$&$1.6\cdot10^{-8}$ &$<0.0003$ & &0.23$^*$&\\
Mars&3397&$2.6\cdot10^{12}$&$2.7\cdot10^{-5}$&$<0.0005$&& &$\sim$1.5$^{**}$\\
Jupiter&71492 &$1.56\cdot10^{20}$&20000&428&-9.6&50-100&\\
Saturn&60268&$4.6\cdot10^{18}$&580&22&0&16-22&\\
Uranus&25559&$3.9\cdot10^{18}$&50&23&-59&$\sim$18-25&$\lesssim33$\\
Neptune&24764&$2.2\cdot10^{17}$&27&1.42&-47&23-26&$\lesssim34$\\ [1ex]
 \hline 
 \multicolumn{8}{l}{} \\[-0.7ex]
 \multicolumn{8}{l}{$^\dag${\footnotesize Variation due to the large Mercury eccentricity. $\quad ^\ddag$Surface field strength.}}\\[-0.3ex]
\multicolumn{8}{l}{$^*${\footnotesize No gobal magnetopause; global magnetic data are upper limits.}}\\[-0.3ex]
\multicolumn{8}{l}{$^{**}${\footnotesize Mars Global Surveyor observations \citep{Acuna2001}.}}\\[-0.3ex]
\multicolumn{8}{l}{{~~~}{\footnotesize Giant planet data from \cite{Bagenal1992}.}}
 \end{tabular}
\end{center}
\vspace{-0.5cm} 
\label{chapPBS-table2}
\end{table}

Such localised fields have been observed only recently to be of the order of a few100 mV/m yielding localised potential wells $\Delta \phi \sim$ kV in the ramp and overshoot region. They are capable of reflecting and accelerating charged particles. A substantial fraction of such localised potential drops of the order of few 10 mV/m is found  already in the foot region, but the main place where they exist is in the shock ramp and overshoot. From simulations, the overshoot is known to be the region where the strongest ion reflection occurs and the upstream ion flow mixes into the downstream hot ions while in the ramp ion phase space vortices develop where ions are locally trapped and which are related to oscillations in the magnetic field with the magnetic maxima occurring at the places of the vertexes of the vortices.  So far the particle observations have not been able to resolve similar structures. These are however expected in the regions of the localised electric fields which in the plasma wave spectra appear as broadband noise reaching from low frequencies near the ion plasma frequency up to well above the electron plasma frequency.

The quasi-parallel bow shock possesses an extended electron foreshock and a broad ion foreshock which are parts of the shock transition. 

The ion foreshock contains upstream ULF waves which at large distance from the shock are quasi-periodic but increase in amplitude when approaching the shock transition. Several types of such waves have been observed. The quasi-periodic waves are mainly Alfv\'en-ion cyclotron waves having upstream directed phase velocity, magnetosonic waves evolve into shocklets, and some of the waves evolve into large amplitude pulsations of irregular shape implying a broadband magnetic spectrum. They are generated when the upstream waves are convected downstream toward the shock and start resonantly interacting with the upstream diffuse ion component that is produced at the shock. Approaching the shock the density of diffuse superthermal ions increases about exponentially causing the interaction to readily become nonlinear, causing the pulsation wave amplitudes to grow and steepen during the downstream convection toward the shock ramp. In fact each pulsation when becoming sufficiently steep and large amplitude starts behaving like a part of the shock, reflecting some ions and generating other waves, e.g. standing whistlers, in front of the pulsation.

The large amplitude pulsations have been found of being the generators of the shock in quasiparallel shock reformation. Quasi-parallel shock reformation works by replacing the shock front with newly arriving large amplitude pulsations. This is a local process since each pulsation is a three-dimensional structure of relatively irregular shape, limited width in the direction normal to the shock and limited extension in both directions along the shock surface. This makes the quasi-parallel bow shock highly irregular and variable in time and space. 

Production of these diffuse superthermal energetic particles and the very mechanism of generating the pulsations is not yet understood satisfactorily. The parallel shock is capable of accelerating ions nonadiabatically, injecting them into the shock environment where they may become the seed population for further acceleration by the first order Fermi process. This might be done by a shock surfing mechanism which requires trapping of the ion component at the shock transition. At the bow shock the threshold energy for ions for stepping into the Fermi acceleration is in the range of $\sim$(50-100) keV/nucleon.
The particle trapping and acceleration probably occurs by trapping of ions between the old shock ramp and the new pulsation. Acceleration of particles takes than place when the trapped ion feel the motional tangential electric field which is continuous across the shock. Probably electrons are accelerated in a similar way since it has been found that the upstream waves and in particular the pulsations with their transverse magnetic fields cause a local turning of the shock normal in the perpendicular direction. The quasi-parallel bow shock probably behaves locally like a quasi-perpendicular shock  on distances $<\lambda_i$ ahead of the nominal shock ramp. 

The electron foreshock on the other hand contains electron beams that are reflected from the tangential part of the shock. These electron beams excite Langmuir waves and cause electromagnetic radiation from the shock. Both has been observed from remote as well as in situ of the bow shock. The Langmuir waves are restricted to the magnetic flux tube that is tangential to the shock surface. They thus indicate that the region of observation is magnetically connected to the shock. the radiation, on the other hand, can be observed from remote. It is usually emitted at the harmonic $\omega\approx 2\omega_pe$ of the local plasma frequency at the location of the excitation of Langmuir waves. It thus carries information about the plasma density upstream of the shock.  

Many of the processes noted here have not yet been fully understood even though the Earth's bow shock is the best investigated shock in the entire solar system. Some of the claims made above are even speculative and are based on the combination of insufficient observations combined with insufficient (one-dimensional or hybrid) simulations which do not yet describe the complex and interwoven processes which act in forming the bow shock. 

\subsection{Mercury's Bow Shock}\index{planets!Mercury (Hermes)}\index{planets!Hermes (Mercury)}
\noindent Mercury, the smallest and closest to Sun planet, is in several respects similar to Earth and different from the other planets. Relative to its size Mercury has the largest iron core and the thinnest mantle of all terrestrial planets. Because of this reason we treat its bow shock at this place first after Earth's bow shock in order to demonstrate the similarities and differences. \index{shocks!Mercury's bow shock}

Mercury might have developed into another Earth if it would not have been so close to the Sun where it was exposed to a continuous bombardment by other heavy celestial bodies that were attracted by the Sun. Its surface speaks of the many collisions it had to survive. However, being on a stable orbit, it did quite well despite that in these collisions Mercury has  been skinned loosing its crust and part of the mantle such that practically only its iron nucleus surrounded by a comparably thin mantle shell was left over. 

Its surface also speaks of the strains Mercury experienced during its evolution and still experiences on its extremely eccentric orbit around the sun which the sun exerts on it, driving tectonic activity on Mercury. Still, Mercury managed to maintain a weak interior magnetic field  the origin of which is not yet clear and gives rise to speculations about a surviving dynamo with contributions that are driven by induction from the outside. The magnetic moment of Mercury is about $<1\%$ of that of the Earth. Mercury's dipole moment is ${\textsf M}_{\rm Merc}=4\times10^{-4}{\textsf M}_{\rm E}=(4.9\pm0.2)\times10^{21}$ nT\,m$^3$ \citep{Ness1978}. These and other questions have been discussed in a recent monograph on Mercury \citep{Balogh2008}. The internal dynamo  Mercury possesses may be the result of surviving tectonic activity or is kept alive by the strains. 
\begin{figure}[t!]
\centerline{\includegraphics[width=1.0\textwidth,clip=]{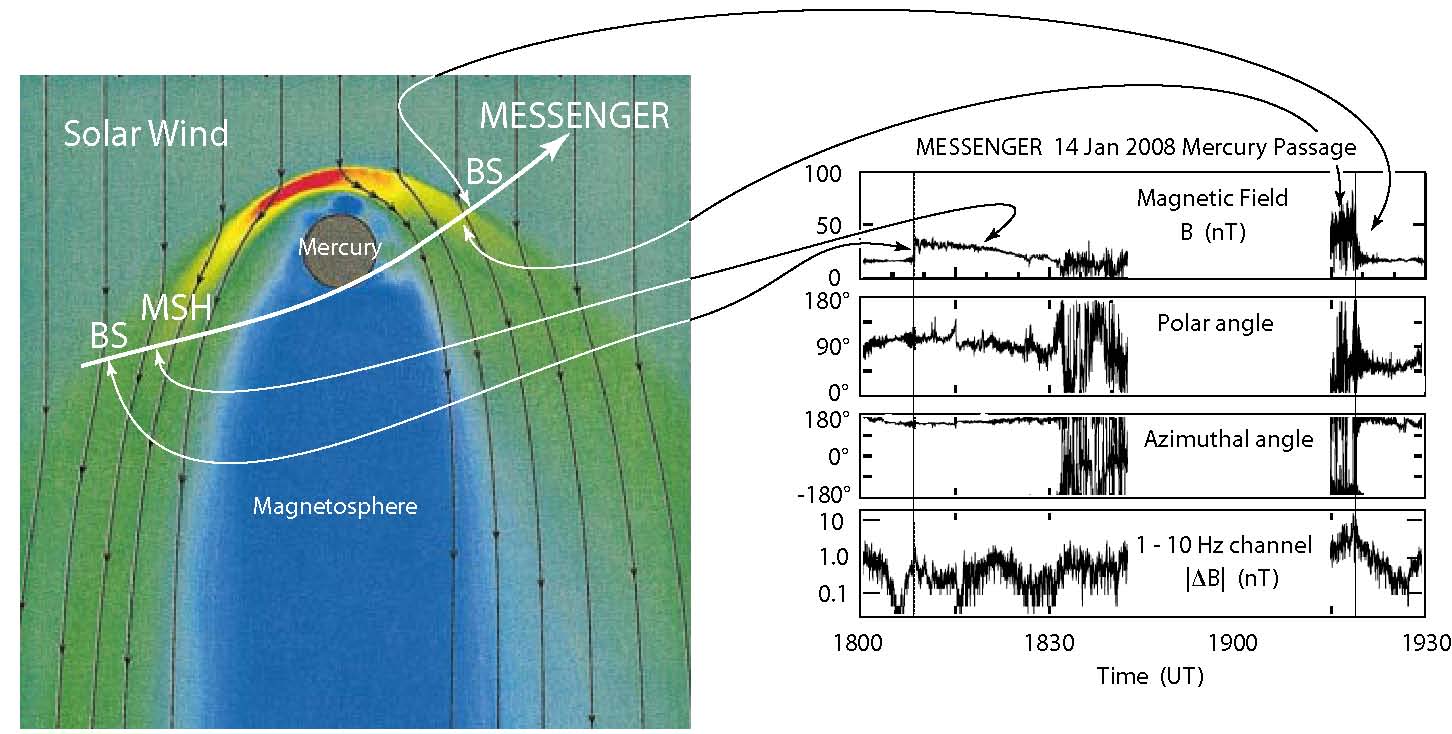}}
\caption[] 
{\footnotesize The MESSENGER magnetic field observations \citep[after][]{Anderson2008} during the Mercury bow shock crossings. Shown is the spacecraft orbit near Mercury overlaid on an MHD simulated Mercury magnetosphere \citep[simulation from][]{Kabin2000}. Only the shock and magnetosheath crossings are shown. The time from 1845 UT to 1915 UT the spacecraft was in the Mercury magnetosphere.\index{magnetosphere!Mercury} The two bow shock crossings are quite different of different compression ratio and dynamics. Note the small foot at the first crossing and extended magnetosheath behind the bow shock. In the second crossing the magntosheath is narrow and very turbulent, the shock has a large overshoot and a much broader foot.}\label{chapBS-fig-Mess1}
\end{figure}

Mercury's magnetic field though weak surrounds the planet and interacts with the solar wind which is quite strong at the Mercury orbit (see Table \ref{chapPBS-table1}). Thus Mercury behaves like a magnetic planet and necessarily develops a bow shock in front of its magnetosphere. So far there have been only three spacecraft encounters with Mercury that provided magnetic measurements near Mercury, two of them encounters of {\small Mariner 10}, and one recent encounter of {\footnotesize MESSENGER}. \index{spacecraft!Mariner 10}\index{spacecraft!MESSENGER} {\small Mariner 10} found the magnetopause solar wind stagnation point (the nose distance of the Mercury magnetosphere) at a planetocentric distance of $R_{\rm MP,Merc}=(1.85\pm0.15)$ R$_{\rm Merc}$ (the Mercury radius being R$_{\rm Merc}=2439\pm1$ km), which is very close to the planetary surface and leaves very little space between the bow shock and magnetopause distances at the nose. Applying the canonical formula Equ. (\ref{chapPBS-eq-BSdist}) gives for this distance just $\Delta_{d, \rm Merc}=1240$ km. (Note also that the determination of the magnetopause and shock distance is quite uncertain, in Table \ref{chapPBS-table2} the values given are smaller even though they are based on the same data of {\small Mariner 10}. This is caused by a high uncertainty of the Mercury magnetic field, the variability of the solar wind, Mercury's orbital eccentricity and the location where the spacecraft crosses the bow shock which implies the assumption of a magnetospheric model.) Compared to the solar wind ion gyroradius of $r_{ci}\sim 300$ km at Mercury this Mercury magnetosheath transition region is only $4r_{ci}$ wide.  {\small Mariner 10} passed the bow shock at the flanks where the distance between  bow shock and magnetopause is large enough to distinguish the two boundaries in the magnetic field recordings. From the behaviour of the magnetic field at bow shock encounter it was concluded that one of the bow shock passages was at a quasi-perpendicular bow shock, the other happened to cross a more quasi-parallel shock \citep{Ogilvie1974,Ness1978}. 

The recent {\footnotesize MESSENGER} passage \citep{Solomon2008,Anderson2008,Slavin2008} gave a similar result. Figure \ref{chapBS-fig-Mess1} shows the magnetic field along the {\footnotesize MESSENGER} path across the Mercury bow shock-magnetosphere system. As before {\small Mariner 10}, the {\footnotesize MESSENGER} spacecraft detected a magnetosphere around Mercury. It crossed the bow shock and magnetosheath twice at different locations. This is illustrated in the figure by the use of a fluid model simulation \citep{Kabin2000}. On its inbound path the spacecraft entered from a quiet solar wind across a nearly perpendicular flank bow shock which exhibited a small foot region in front of a weak shock of compression ratio $<2$, afterwards staying in a broad flank magnetosheath. The sheath region adjacent to the shock was quite calm, large amplitude oscillations in magnetic field and direction occurred only close to the magnetopause encounter and are probably standing mirror modes known from Earth's magnetosheath. The fluctuations are seen in the polar and azimuthal angles of the magnetic field.  The amplitude of low frequency fluctuations $<10$ Hz in the magnetic field are plotted in the lowest panel. Interestingly the shock crossing does not occur as a spectacular event in these fluctuations. 

The outbound bow shock crossing exhibits a highly fluctuating magnetosheath and another sharp much stronger shock crossing of compression factor $\sim 3$ and a large overshoot. There are strong fluctuations in front of this shock crossing and a more disturbed solar wind following the crossing. However, the nature of this crossing is also more quasi-perpendicular than quasi-parallel with an extended active shock foot region in front of the shock ramp. The angular variation in the magnetic field shows that the shock angle has varied very strongly during this crossing. The bow shock at this location seemed very variable indeed. Particle observations indicate that during this crossing energetic ions were present in the immediate shock environment \citep{Zurbuchen2008}. Taken these observations together with the {\small Mariner 10} observations one concludes that Mercury possesses a supercritical bow shock that is continuously present and is a moderately strong shock. It develops a foot and a highly turbulent downstream magnetosheath and has regions of different strength. Even though this {\footnotesize MESSENGER} passage gave no indication for a parallel shock region, {\small Mariner  10} had crossed a quasi-parallel shock.

All this is very similar to Earth's bow shock though on a smaller scale and in a different environment. However, a difference between Earth's and Mercury's bow shocks lies in the fact that Mercury has been found to be surrounded by a comet-like cloud of ions. These ions are of planetary origin, probably being created by sputtering from the Mercury surface \citep{Slavin2008}. Since it is known that such sputtered ions when entering the solar wind become pickup ions and are accelerated, they could considerably affect the nature of the Mercury bow shock by reflection and acceleration. Thus Mercury's bow shock is in some sense a hybrid bow shock somewhere in between a comet and a magnetised planet. The implications of this finding for the structure and dynamics of the Mercury bow shock are not yet known.

\subsection{Jupiter's Bow Shock}\index{planets!Jupiter (Jova)}\index{planets!Jova (Jupiter)}
\noindent Jupiter is not only the largest planet, it has also the strongest magnetic field believed to be generated by a huge internal dynamo that is driven by differential rotation in the planet interior. Indeed Jupiters rotation period is short, less than half as long as a terrestrial day. Table \ref{chapPBS-table2} shows the surface strength of this field to be more than an order of magnitude larger than the Earth's surface field. Together with the decreased kinetic solar wind pressure at the 5.2 AU heliocentric distance of Jupiter this strong field promises Jupiter to be surrounded by a huge magnetosphere that forms a giant obstacle in the solar wind and causes an extended bow shock to stand in front of the magnetosphere.\index{magnetosphere!Jupiter} \index{shocks!Jupiter's bow shock}

Simple scaling from Earth yields a stand-off distance for the magnetospheric nose of 4.2 times the Earth's magnetopause distance measured in Jovian radii. This yields a theoretical value of $R_{\rm MP, J}= 46\, {\rm R_J}$, a value that underestimates the observed distance which varies between 50 ${\rm R_J}$ and 100 ${\rm R_J}$. This variation is due to the internal magnetospheric dynamics of the planet which is determined by the fast planetary rotation which exerts enormous centrifugal forces acting on the cold plasma in the inner magnetosphere, the Jovian plasmasphere. These forces cause a strong deviation of the plasmasphere stretching the plasma radially out to become a fat sheet that extends out to the magnetopause, carrying the Jovian magnetic field with it and increase both the plasma and magnetic pressure at the contact with the solar wind. Since this happens in the symmetry plane of the Jovian rotation the nose of the Jovian magnetosphere deviates from bluntness and becomes more cusp-like than the Earth's magnetosphere.  
\begin{figure}[t!]
\centerline{\includegraphics[width=1.0\textwidth,clip=]{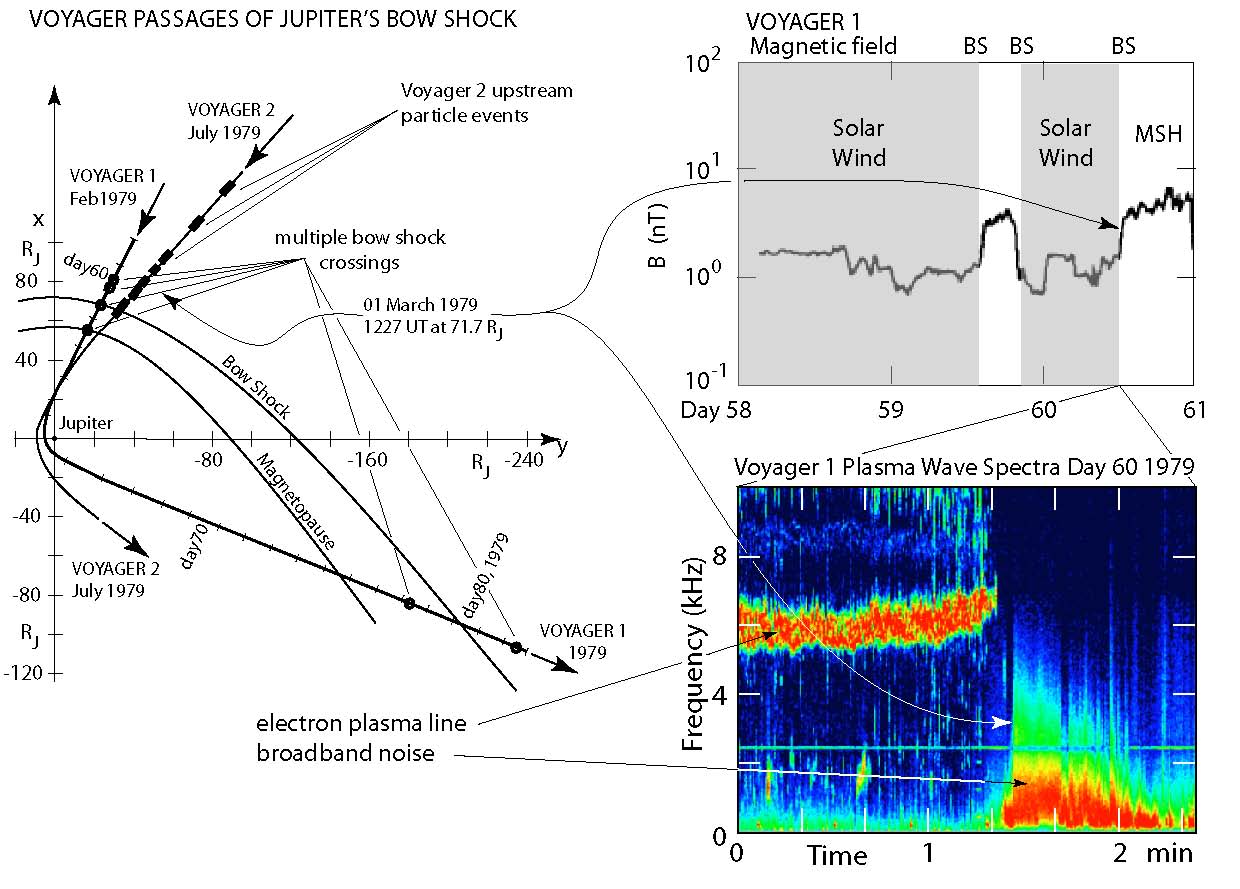}}
\caption[Saturn] 
{\footnotesize Summary of Voyager 1 crossings of the Jovian bow shock. {\it Left}: Projected Voyager  1 orbit in 1979 around Jupiter and the nominal bow shock and magnetopause locations in the $(x,y)$-plane with $x$ pointing towards the sun. The numbers along the Jupiter path give the days of the year 1979. The dots along the orbit indicate the multiple crossing of the Jupiter bow shock showing its high variability \citep[after][]{Ness1979}. Also shown is part of the Voyager 2 orbit in July 1979. The black bars on this orbit indicate the orbital segments when energetic particle events have been detected upstream of the bow shock \citep[after][]{Krimigis1985}. {\it Right}: Magnetic field (top) and plasma wave spectra (bottom) during selected crossings. The magnetic field shows the increase (decrease) when Voyager passes the shock inbound (outbound). Two minutes of plasma wave spectra are shown below for the third crossing in the top figure (red is highest, black lowest wave intensity). Voyager is in the electron foreshock before crossing as seen from the intense plasma frequency at 6 kHz. Passage of the shock occurs as a sharp cutoff of the plasma frequency and the start of broadband low frequency noise (courtesy by D. A. Gurnett, U Iowa). }\label{chapBS-fig-JovaVoy1}
\end{figure}

At the Jovian orbit the Mach numbers are large, ${\cal M}>10$ even under undisturbed conditions. The huge obstacle of Jovian magnetosphere thus causes a huge bow shock to stand in front of it. This bow shock is highly supercritical and should have a high compression ratio $N_2/N_1\lesssim 4$.  Straightforward application of Eq. (\ref{chapPBS-equ-Deltad}) gives for the distance between the Jovian nose magnetopause and the Jovian bow shock the interval $13\,{\rm R_J}\lesssim \Delta_{d{\rm J}}= 0.275\, R_{\rm MP, J}\lesssim 28\,{\rm R_J}$. This is a substantial distance which, even in terms of numbers of planetary radii, is larger than the geocentric distance of Earth's bow shock from Earth! Since the solar wind magnetic field at the Jupiter orbit is of the order of $B\sim 1$ nT, the 1 keV ion gyroradius becomes $r_{ci}\approx 750$ km, vanishingly small compared with the distance between the bow shock and magnetosphere. Moreover, the solar wind number density at the Jovian orbit has decreased from that at Earth's orbit by a factor of 27. The ion inertial length becomes $\lambda_i(5.2\,{\rm AU})\approx 530$ km. The Jovian bow shock is thus extremely thin, measured in Jovian radii. Everything happens on the plasma scales and not on the scales of the obstacle which only determine the gross scales like stand-off distance and curvature radius. 

One therefore expects that, up to the lesser plasma densities and magnetic fields, the Jovian bow shock behaves very similarly to Earth's bow shock. No large surprises are expected in crossing it except for the slightly different plasma properties like lower densities, temperatures and magnetic fields as well as higher kinetic $\beta$ corresponding to higher Mach numbers. This implies a somewhat stronger shock, stronger ion reflection, larger two stream and modified two stream instability growth rates and, possibly, a better efficiency in particle acceleration.  In addition, due to its considerably larger curvature radius, the quasi-perpendicular and quasi-parallel surface areas on the Jovian bow shock are much larger than in the case of Earth's bow shock. This larger area combined with the much wider magnetosheath region will support particle acceleration simply because the foreshock region is more extended and the downstream magnetosheath is broad enough for letting the turbulence evolve into a state closer to the state of well developed turbulence than in the case of Earth's magnetosheath. 

The above values are the theoretical expectations which are based on the knowledge of Earth's bow shock. Jupiters bow shock has been crossed by a small number of spacecraft, {\small Pioneer  10}, {\small Pioneer11}, the two {\small Voyager} spacecraft, the {\small Galileo} mission, and {\small Ulysses}. Already the two {\small Pioneer} spacecraft confirmed the existence of a bow shock at Jupiter. {\small Pioneer 10} observed upstream waves in association with bursts of relativistic electrons from Jupiter. {\small Pioneer 11} crossed the Jupiter bow shock six times at planetocentric distances in the interval $78\leq R_{\rm BS,J}\leq 110$ R$_{\rm J}$. The average shock velocity in the solar wind frame was found to be of the order of $V_{\rm BS}\sim 100$ km/s. A turbulent magnetosheath was detected where the average magnetic field was draped around the Jupiter magnetosphere in a way suggesting a blunt magnetopause shape. Moreover, large amplitude waves were reported to be observed in the magnetosheath \citep{Smith1975}. 

In 1979 the two {\small Voyager} \index{spacecraft!Voyager} spacecraft passed Jupiter on their way out into the outer heliosphere and interstellar space. During this passage which preceded the passages of the other giant planets of the outer solar planetary system they also crossed Jupiter's bow shock when entering close to the planet in a swing by maneuvre. Figure \ref{chapBS-fig-JovaVoy1} shows the path of {\small Voyager l 1}  near Jupiter in February 1979. Again the bow shock was crossed six times (indicated by circles on the spacecraft path). Each time it was crossed multiply, indicating the very high variability of the Jovian bow shock location. The magnetic recording of three of the crossings (on days 59 and 60, 1979) is plotted in the upper right panel of the figure. All three crossings belong to the class of quasi-perpendicular shocks exhibiting steep compressions of the magnetic field and some indication of a foot in front of the shock ramp, The compression is about a factor 3 identifying the Jovian bow shock as a fairly strong but not a superstrong shock. Two minutes of plasma wave spectrogram during the shock crossing \citep[see][]{Scarf1979,Scarf1979a} are given on the lower right. The spectrogram covers the range of frequencies $<10$ kHz which includes the nominal plasma frequency based on the expanding solar wind model. Indeed, prior to the shock crossing an intense plasma frequency line at $\sim 6$ kHz was detected that is seen here as a red emission band which disappears roughly 5 s  before the spacecraft enters the shock. Approaching the shock the frequency increases gradually thereby mapping the increase in plasma density. Taking the nominal shock velocity of 100 km/s the 5 s between the end of the plasma line and the shock crossing correspond to a distance of 500 km upstream of the shock, roughly comparable to the upstream ion gyroradius and the width of the shock foot. Indeed, some weak broadband noise begins at this time filling the gap from below. Referring to our knowledge about the processes taking place in the quasi-perpendicular shock foot we may conclude that these are the waves which are generated by two-stream and modified two-stream instabilities in the shock foot. Finally, when entering the shock ramp the broadband noise suddenly intensifies at the low frequencies and the bandwidth increases above the pre-shock plasma frequency. This is the typical signature of a quasi-perpendicular shock ramp crossing. For a shock of compression factor 3 the ramp plasma frequency is above 10 kHz and thus not captured in this spectrogram. Similarly, any radiation that is possibly emitted at the harmonic of the plasma frequency in front of the shock is at $\sim$12 kHz outside the spectrogram. There is indication of a weak emission line at 8-9 kHz the origin of which is not known. More than a decade later, in February 1992 the {\small Ulysses} spacecraft on its swing-by path around Jupiter observed similar plasma wave emissions from the Jovian bow shock \citep{Stone1992}.

Similar to {\small Voyager  1}, {\small Voyager  2} who crossed Jupiter's bow shock in July 1979 observed sporadic energetic electron bursts related to electron plasma oscillations and ion acoustic waves for several weeks  prior to crossing the shock \citep{Gurnett1979}. In analogy to Earth's bow shock these observations suggested magnetic connection to Jupiter's bow shock. When approaching the planet, energetic ion bursts were detected  within the energy interval $30\,{\rm keV}\lesssim {\cal E}_i\lesssim 5\,{\rm MeV}$ \citep{Krimigis1985}. The intervals when this happened are shown in Figure \ref{chapBS-fig-JovaVoy1} as black bars along the {\small Voyager 2} orbit. The energy spectra of these particles in the energy range $0.1\,{\rm MeV}\lesssim {\cal E}_i\lesssim 2$ MeV followed a power law shape in energy. Above 300 keV the spectra were void of protons being dominated by Oxygen and Sulphur, which indicates that Jovian magnetospheric ions were leaking out and were accelerated by the Jovian bow shock.

In toto {\small Voyager 2} experienced  five multiple crossings of Jupiter's bow shock between 99 R$_{\rm J}$ and 62 R$_{\rm J}$, in good agreement with the {\small Pioneer  11} observations of the location of Jupiter's bow shock. On 5 July 1979 {\small Voyager  2} reported a multiple crossing at planetocentric distances between $R\sim$69 R$_{\rm J}$  and $R\sim$64 R$_{\rm J}$ in a time interval $\sim 5$ hrs long, witnessing very high variability of the location of the bow shock. Since the Jupiter day is only 9.9 hrs long, the planet performed half a full rotation during this time such that part of the bow shock variability was caused by the wobbling of Jova's magnetosphere and the flapping of its  dayside plasmaspheric nose. One may conclude from this that Jupiters bow shock position and form are highly variable being affected not only by the conditions in the solar wind but also by the Jovian magnetosphere in a way that is much stronger than Earth's bow shock is affected by Earth's magnetosphere. 
\begin{figure}[t!]
\centerline{\includegraphics[width=1.0\textwidth,clip=]{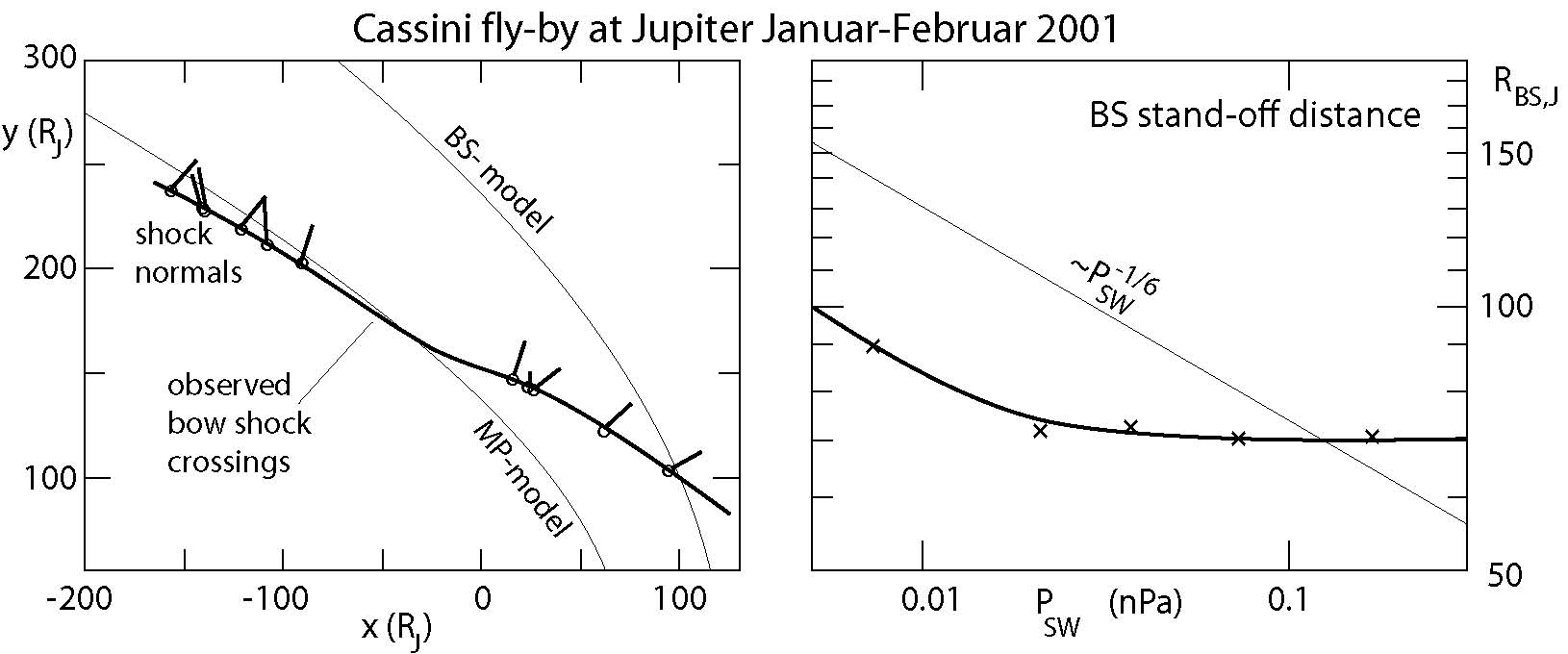}}
\caption[Saturn] 
{\footnotesize  Jupiter's  bow shock location as inferred from crossings of the Cassini spacecraft during its fly-by at Jupiter on its way to Saturn. {\it Left}: The model positions of the Jovian bow shock and magnetopause and the observed locations of the bow shock showing including the shock normal directions. One notes the variations in the shock normal which indicate the non-stationary conditions at the bow shock. {\it Right}: The inferred dependence of the stand-off distance of the Jovian bow shock as function of the solar wind pressure. The straight line shows the predicted theoretical dependence. The observation indicates independence of the stand-off distance from solar wind pressure for pressures exceeding 0.02 nPa  \citep[data taken from][]{Achilleos2004}. }\label{chapBS-fig-Achilles}
\end{figure}

The Jovian bow shock has also been crossed by the {\small Ulysses} spacecraft in February 1992 when it passed Jupiter. In the late nineties the {\small G}alileo spacecraft crossed it several times, and {\small Cassini}, during its Jupiter flyby on the way to Saturn skimmed the Jovian bow shock at the time when {\small Galileo} was also in orbit. \cite{Achilleos2004} analysed the magnetic field of 30 passages of the more than 40  {\small Cassini} spacecraft crossings of Jupiter's bow shock between October 2000 and April 2001. Their important finding is that the stand-off distance of the Jovian bow shock is relatively independent of solar wind pressure for pressures above ${\textsf P}_{\rm SW}\gtrsim0.02$ nPa. At these pressures the average shock nose distance was around $\sim$70 R$_{\rm J}$. At lower solar wind pressures the stand-off distance of the shock increased. This is interesting as it suggests that the Jovian magnetosphere becomes incompressible when it is compressed to a distance of about $\sim 70$ R$_{\rm J}$, an effect which may be due to the presence of the extended internal magnetospheric corotating Jovian plasma disk which adds to the resistance of the magnetosphere against further compression. At solar wind pressures ${\textsf P}_{\rm SW}\lesssim0.02$ nPa the magnetosphere is sufficiently expanded to follow the theoretical law $R_{\rm BS,J}\propto {\textsf P}_{\rm SW}^{-1/6}$ given by \cite{Slavin1985}. This is also concuded from the determined shape of the bow shock on the left in Figure \ref{chapBS-fig-Achilles} which shows that the stand-off distance of the bow shock in the ecliptic is farther out than at higher latitudes. At higher latitudes the effect of the plasma disk vanishes and the shock position is supported only by the Jovian magnetic field in the magnetosphere. However, these single spacecraft measurements are ambiguous leaving it open whether this ecliptic expansion was not a temporal effect caused by a change in solar wind pressure. Such a change has indeed been observed simultaneously by {\small Cassini} and {\small Galileo} \citep{Kurth2002} and has been interpreted as deformation of the magnetopause.

The second interesting result of this investigation is that the shock normal exhibits substantial variations. The Jovian bow shock is highly variable in time, which is not an unexpected result seen in Figure \ref{chapBS-fig-Achilles}.  The farthest {\small Cassini} bow shock crossing where plasma wave observations were available occurred at planetocentric distance of 140.2 R$_{\rm J}$ again confirming the existence of Langmuir waves and ion acoustic waves in the foreshock, and an intense burst of broadband emissions during the shock ramp passage \citep{Kurth2002}. 12 of these crossings were well-defined quasi-perpendicular shocks, 5 were quasi-parallel, the rest were not well defined oblique shocks. {\small Cassini} found the bow shock to extend over 700 R$_{\rm J}$ down the flank of Jupiter's magnetosphere \citep{Szego2003}. Plasma observation during these crossing showed that the shock crossings were characterised by the expected strong ion thermalisation which should coincide with reflection of ions and the appearance of an electrostatic shock potential. Both effects gradually weaken with increasing distance from the bow shock nose. In particular the shock potential softens rapidly, indicating a weaker and more variable shock at increasing downstream distance. \cite{Krupp2004} refer to energetic particle observations during {\small Cassini} bow shock crossings of the far downstream bow shock which show the occurrence of energetic ions $>36$ keV and in a few cases also relativistic electrons. Though the ions are clearly related to the existence of the shock probably being of diffusive origin, the relativistic electrons are not attributed by these authors to acceleration by the Jovian bow shock; rather they believe they leak out from the Jovian magnetosphere due to reconnection. Thus the picture is rather confused as electron acceleration can also be achieved by trapping in magnetic flux tubes that are connected to the shock, which is not unreasonable in view of the huge extension of Jupiter's bow shock.

\subsection{Saturn, Uranus, Neptune}\index{planets!Saturn}\index{planets!Uranus}\index{planets!Neptune}
\noindent The remaining three giant outer planets are all strongly magnetised (see Table \ref{chapPBS-table2}). As a consequence they possess magnetospheres. Since the solar wind dynamic pressure is relatively low at their locations, these magnetospheres are very large, and strong bow shocks at large stand-off distances from the planets evolve in front of them. Moreover, their relatively fast spin periods (Table \ref{chapPBS-table1}) together with their large radii suggest that their magnetospheres are also subject to strong corotation of a presumable inner magnetospheric plasma component. Whether such a component exists is a question as the solar UV radiation turns out to be rather weak at the distances of these planets. Nevertheless, cosmic ray produced ionospheres due to ionisation of their atmospheres will release sufficient plasma along the magnetic field to populate the inner corotating regions of the magnetospheres. In the absence of such an ionosphere the planetary magnetic fields are still strong enough for producing a magnetosphere and its shielding bow shock. 

\subsubsection{The Saturnian bow shock}\index{shocks!Saturn's bow shock}
\noindent The bow shock of the next closest of the outer giant planets, Saturn, was first encountered by the {\small Pioneer  11} spacecraft in 1979  \citep{Smith1980} at a distance of 23.7 R$_{\rm S}$. When {\small Pioneer 11} left the planet,  having crossed its magnetosphere,\index{magnetosphere!Saturn} monitoring the Saturnian magnetic field dipole and confirming the existence of a corotating plasma between $4\,{\rm R_S}\lesssim R\lesssim 16\,{\rm R_S}$, it again experienced multiple bow shock crossing between distances of $49.4\,{\rm R_S}\leq R\leq 65\,{\rm R_S}$. Similar to Jupiter, the corotating plasma component (called Saturn's plasma torus) plays some role in the dynamics of the Saturnian magnetosphere. Its particle energies are in the range $0.1\lesssim {\cal E}\lesssim 8$ keV, and plasma densities are of the order of $N\sim5\times10^5\,{\rm m}^{-3}$, and temperatures $T\sim 10^6$ K. Thus the plasma-$\beta\sim 1$ up to an inward distance of $\sim 6.5\,{\rm R_S}$. Probably, this plasma is caused by photodissociation of water frost of Saturn's rings \citep{Frank1980}. 

One year later {\small Voyager 1} and then {\small Voyager 2} passed Saturn in November 1980 and August 1981, respectively, crossing its bow shock and magnetosphere and confirming each others observations. Altogether the observations of these spacecraft identified Saturn's bow shock to be very similar to Jupiter's bow shock. The stand-off distances have already been given in Table \ref{chapPBS-table2} showing that gain as in the case of Jupiter the Saturnian bow shock is far upstream in absolute units and thus very thin being of the plasma scale and leaving a broad (in terms of plasma scales) magnetosheath downstream behind it. At Saturn's orbit the solar wind is even colder ($T_i\sim 1.4$ eV, $T_e\sim 2$ eV) and more dilute than at Jupiter with $N\sim 1.1\times10^5$ m$^{-3}$, the Alfv\'en and magnetosonic speed are low, $V_A\sim 20$ km/s and $V_{ms}\sim 28$ km/s, respectively \citep{Scarf1981}. Thus the Mach numbers are high, of the order of ${\cal M}\sim 15-20$, and the shock is strong. From {\small Voyager 1} an inbound dayside high Mach number quasi-perpendicular shock crossing at planetocentric distance of 32.4 R$_{\rm S}$ and an outbound morning flank quasi-parallel shock crossing at 78 R$_{\rm S}$ was reported, the latter possessing an extended turbulent foreshock region containing much low-frequency upstream turbulence \citep{Ness1981} 

Plasma wave observations from both {\small Voyager} spacecraft upstream of the shock confirmed occasional magnetic connectivity from Langmuir and ion acoustic waves and suggesting the presence of fast electron beams. Just prior to crossing the bow shock strong Langmuir wave bursts were observed, and the shock manifested itself in broadband low frequency noise as was expected \citep{Gurnett1981,Scarf1982}. One surprising observation of {\small Voyager 1}was that downstream of the shock the plasma wave intensities were rather low, much less than in Earth's magnetosheath, and observation that still awaits explanation. \cite{Scarf1981}, from comparison of Earth, Jupiter and Saturn plasma wave spectra, speculates that there is a relation between Mach number and wave excitation. This is indeed the case as at high Mach numbers other instabilities are excited. However, the downstream absence of high wave levels is more difficult to understand. Rather it has to do with the enormous downstream extension of the giant planet magnetosheaths where the high levels of turbulence are confined to the downstream region adjacent to the bow shock with thickness of the order of a couple of energetic passing ion gyroradii. Farther downstream outside the thermalisation distance the plasma probably forgets for the presence of the shock and normalises at a submagnetosonic flow state of enhanced temperature and density. On the other hand, {\small Voyager 2} observed very low plasma densities and plasma frequencies at the bow shock crossings. From bow shock models these low values seemed unusual, and it was suggested that the wake of Jupiter's extended magnetotail swept over Saturn during the passage of {\small Voyager 2}.

\begin{figure}[t!]
\vspace{-0.4cm}\centerline{\includegraphics[width=1.0\textwidth,clip=]{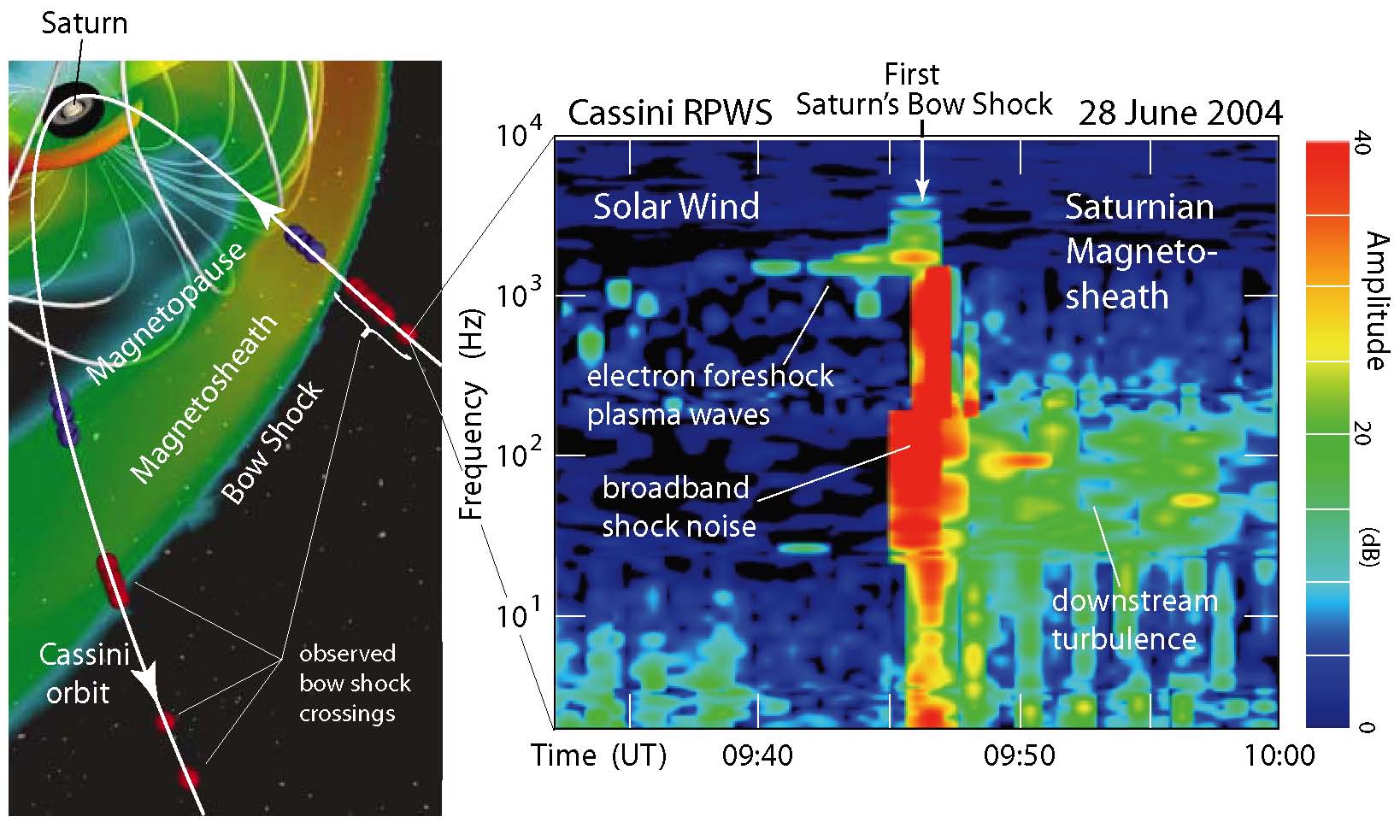}}
\caption[Saturn] 
{\footnotesize The first Cassini spacecraft crossing of Saturn's bow shock at a distance of 49.2 R$_{\rm S}$ on 28 June 2004 as seen in plasma wave spectra (courtesy by D. A. Gurnett, U Iowa) on the right. The crossing occurs at about 0945 UT. About 5 min prior to the shock the spacecraft detects electron plasma waves which are an indication of magnetic connection between spacecraft and shock. As expected, the shock is seen as a sudden spectral broadening from low frequencies to frequencies well above the plasma frequency. The crossing lasts for $\sim$5 min. Afterwards the spacecraft is in the turbulent Saturnian magnetosheath. The picture on the left shows a 3d-MHD simulation of Saturn's magnetosphere using nominal solar wind values at 9.6 AU \citep[section taken from][]{Gombosi2005}. Overlaid is the inbound-outbound Cassini orbit around Saturn. The red dots indicate observed bow shock crossings, to the first of which belong the plasma wave spectra on the right. Blue dots are magnetopause crossings. Note the large inflation of Saturn's bow shock-magnetosphere system compared with the nominal simulation.}\label{chapBS-fig-CassSatPW}
\end{figure}

Two decades later it was the {\small Cassini} \index{spacecraft!Cassini} spacecraft on its way to Saturn's moon Titan \index{moons!Titan}which crossed the bow shock of Saturn 17 times. {\small Cassini}'s first contact with Saturn's bow shock was on 22 March 2004, when it was hit by electrons beams arriving from the direction of Saturn along the interplanetary spiral magnetic field. These electrons indicated magnetic connectivity to the bow shock at an upstream distance of 825 R$_{\rm S}$ and made themselves visible by exciting a burst of Langmuir waves \citep{Gurnett2005}. Such bursts have been detected in the following during the {\small Cassini} approach more frequently until the first crossing of the Saturnian bow shock on 28 June 2004 at 0945 UT which happened at radial planetocentric distance 49.2 R$_{\rm S}$. Figure \ref{chapBS-fig-CassSatPW} on the right shows the dynamic plasma wave spectrum of this first Saturnian bow shock crossing in the RPWS recordings between 3 Hz and 10 kHz. In the {\small Cassini} plasma instrument RPWS data the crossing appeares as the abrupt transition from the Saturnian electron foreshock plasma oscillations at frequency $f\sim 1.78$ kHz to a burst of low frequency broadband electrostatic noise. The Langmuir electron plasma oscillations correspond to a plasma density $N\sim 4\times10^4\,{\rm m}^{-3}$ in the dilute upstream solar wind plasma, in very good agreement with the $r^{-2}$-decrease of the solar wind density from 1 AU to 9.6 AU. In the time following this crossing the shock position oscillated six times back and forth across the {\small Cassini} spacecraft. The left part of the figure shows the {\small Cassini} orbit around Saturn overlaid on a 3d-MHD simulation study of the solar wind-Saturnian magnetosphere interaction \citep{Gombosi2005} using the nominal solar wind data at the Saturn orbit. The red dots along the path indicate the various bow shock crossings observed by {\small Cassini}. Blue dots indicate magnetopause crossings. It is interesting to note that the simulation grossly underestimates the inflations of both the Saturnian magnetosphere and bow shock.

In the plasma wave spectra the Saturnian bow shock is preceded by $\sim$5 min of plasma oscillations, the trace of which broadens in both directions to higher and lower frequencies just before shock crossing, an effect known from Earth's bow shock and probably related to the presence of the shock foot and the excitation of either electron acoustic, Buneman or modified two-stream instabilities. This is followed by the intense broadband noise burst of the shock transition with maximum intensity around $\sim$(0.1-1) kHz, one minute later extending up to $\sim$5 kHz. The whole transition lasts for just $\lesssim$3 min. After this time {\small Cassini} is in the Saturnian magnetosheath which in the plasma wave spectrum appears as broadband ion acoustic noise $\lesssim$30 Hz that is modulated by the spacecraft spin, and by about isotropic electric noise between 30 Hz and 300 Hz, indicating that {\small Cassini} entered the region of downstream magnetosheath plasma turbulence. In the plasma waves this behaviour is typical for a quasi-perpendicular shock crossings as is known from crossings of Earth's bow shock wave. Taking the view that the shock is strong and of maximum compression, the density and magnetic fields should be compressed by at most a factor 4, implying a shock plasma frequency $\lesssim$ 4 kHz. Clearly the broadband noise detected during the shock crossing extends beyond this value up to $\sim 6$ kHz or more. Hence, it cannot be produced by a plasma instability. Rather it is the result of crossing spatially localised electrostatic structures of central frequency around 100-1000 Hz. 

Little is known about the shock velocity. The electron and ion inertial lengths are in the respective intervals 1.5 km $\lesssim\lambda_e\lesssim$ 3 km and 700 km$\lesssim \lambda_i\lesssim$ 1300 km. If we assume that the width of the shock transition is of the order of the ion inertial length, then the 2-3 min crossing time of the shock correspond to the spacecraft velocity, suggesting that the shock was about stationary during this first crossing, possibly standing at its maximum possible inflation distance. Moreover, assuming that the scales of the localised small scale structures responsible for the broadband noise are of the order of the electron inertial length, then the bandwidth of the noise suggests that these structures move at velocities of the order of $\sim$1000 km/s.  This speed should be less than the electron thermal velocity, suggesting that the electron temperature in the shock is at least of the order of $T_e\gtrsim 50$ eV. When referring to the upstream low electron temperature in the solar wind which, for adiabatic expansion, is of the order of $T_e\sim$(0.5-1) eV, this implies strong electron heating in the shock transition.
\begin{figure}[t!]
\vspace{-0.4cm}\centerline{\includegraphics[width=1.0\textwidth,clip=]{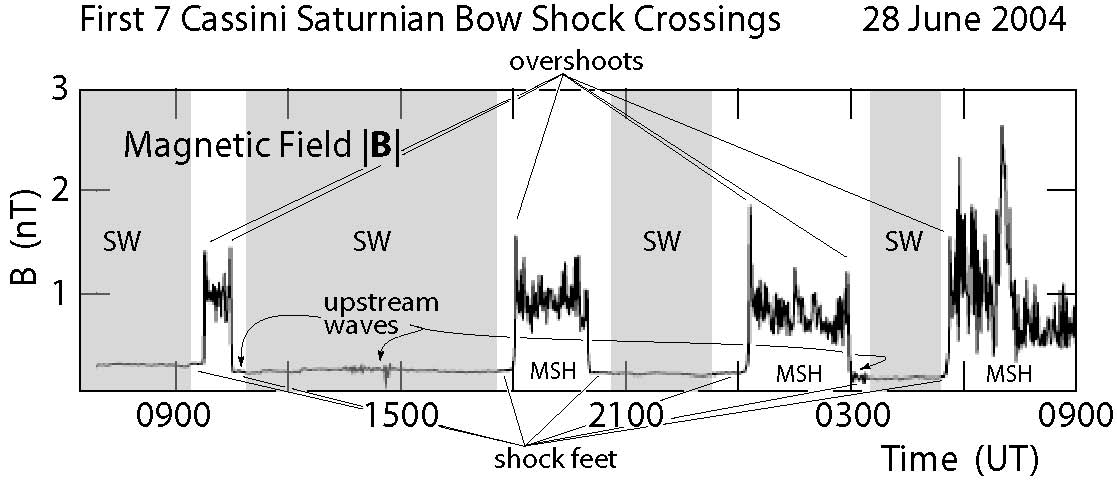}}
\caption[Saturn] 
{\footnotesize The first 7 Cassini spacecraft crossing of Saturn's bow shock \citep[data taken from][]{Achilleos2006} in the magnitude of the magnetic field. All of these crossings are quasi-perpendicular shock crossings with steep ramps, overshoots, feet and indications of ULF upstream waves in their feet. The magnetosheaths are fairly disturbed in the ULF wave range. In one case ULF upstream waves not related to a shock crossing are observed which might either indicate switching into the ion foreshock of a quasi-parallel part of the bow shock or approach to a shock without touching it. The first of these Cassini inbound crossings at 0945 UT is the one shown in the plasma wave spectra in Figure \ref{chapBS-fig-CassSatPW}.}\label{chapBS-fig-CassB}
\end{figure}

The larger than expected inflation of the Saturnian magnetosphere and bow shock points on a more Jupiter-like than Earth-like magnetosphere of Saturn. This is confirmed by a comparative statistical study \citep{Hendricks2005} using data from {\small Pioneer 11}, the two {\small Voyager} spacecraft, and {\small Ulysses}. 

Another question concerns the nature of the shock. As in the case of Jupiter, the bow shock of Saturn shows, in the overwhelming number of cases, a sharp transition from solar wind to the magnetosheath which is typical for quasi-perpendicular shocks. This is seen in Figure \ref{chapBS-fig-CassB} for the first 7 {\small Cassini} bow shock crossings. The absolute changes in the shock ramp are not very high; they are, however, fairly steep with well expressed overshoots. They also possess extended shock feet region which contain ULF wave activity. The very first of these crossings is the same as that shown in Figure \ref{chapBS-fig-CassSatPW} in the high frequency plasma waves. Interesting enough, these spectra indicate that the shock foot is a region of reduced ion acoustic wave activity, which may be in favour of the modified two-stream instability which has lower frequency and is oblique such that convective Doppler shift affects it very little. Since the electron cyclotron frequency is $<10$ Hz, these wave which propagate on the whistler branch are not seen in the spectra and do not occur either in the rough magnetic measurements. \cite{Achilleos2006} have determined the shock strengths. These turn out to be considerably weaker than expected, mostly with compression ratios $B_2/B_1\lesssim2$. Only in one case this ratio is closer to 3. On the other hand, the measured overshoots have been found in the interval $2<B_{\rm over}/B_1<3.5$. Shock ramp widths have been determined to fall between (0.13-1.11)$\lambda_i$ with the sixth crossing making an exception of having a ramp width of $\sim2.75\lambda_i$. Neglecting this crossing, the average ramp width turns out to be $0.6\lambda_i$ which is in agreement of what is known from Earth's bow shock. Shock velocities have been found from passing the shock ramp to be in the average $V_{\rm S}\sim 124\pm108$ km/s. There is a discrepancy in the case of the first crossing where the velocity found is 363 km/s, much larger than the shock velocity determined from the plasma wave spectrum which was between 10-20 km/s. This might result from the fact that the plasma wave spectra in the Figure \ref{chapBS-fig-CassSatPW} do not resolve the shock ramp and thus cover a wider spatial range that includes the overshoot and therefore underestimates the velocity.
\begin{figure}[t!]
\vspace{-0.4cm}\centerline{\includegraphics[width=0.8\textwidth,clip=]{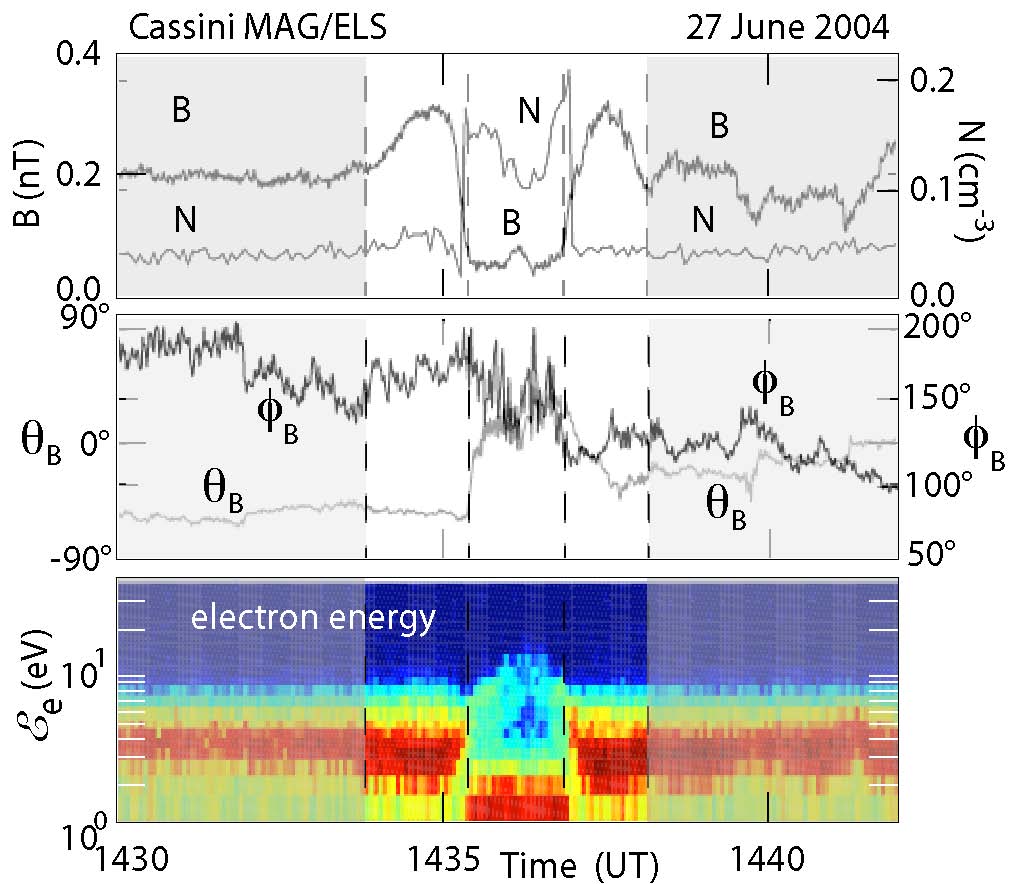}}
\caption[Saturn] 
{\footnotesize Passage of Cassini across an Hot Flow Anomaly near the Saturnian bow shock but before getting into contact with the shock  \citep[data taken from][]{Masters2008}. Shown is the variation in the magnetic field $B$ and density $N$. The Magnetic field develops compressions at the edges of the HFA prior to dropping to very low values inside. Simultaneously the plasma density increases steeply inside the magnetic cavity. The turn in the direction of the magnetic field v ector across the HFA is seen in the two magnetic angles in the middle panel. The bottom panel shows the electron energy colour coded from the solar wind across the HFA back into the solar wind. The electrons are retarded in the boundaries and heated. Inside the HFA the the electrons dont move anymore while assuming a high velocity spread, i.e. a high temperature roughly a factor 1.5 higher than outside.}\label{chapBS-fig-CassHFA}
\end{figure}

The more perpendicular than parallel behaviour of the outer planetary bow shocks is probably related to the more circular nature of the spiral interplanetary magnetic field with increasing heliocentric distance. In the ecliptic plane, the plane of the {\small Cassini}  orbit, the interplanetary magnetic field at the orbit of Saturn is almost azimuthal. Moreover, since the Mach numbers increase with heliocentric distance and the dependence of the velocity-shock normal angle $\theta_{Vn}<\cos^{-1}({\cal M}^{-1})$, the opening angle of the bow shocks for larger Mach numbers increases thus favouring quasi-perpendicular shock formation.  However, this might not always be the case as one particular recent observation shows \citep{Masters2008} where Saturn obviously was crossed by  current sheets of the character of  tangential discontinuities in the solar wind, across that the magnetic field direction rotates by an arbitrary angle possibly also changing magnitude. As a consequence the convection electric field normal to the discontinuity surface may also change sign, depending on the angle of rotation of the magnetic field. If this happens in a way that the electric field to both sides points into the tangential discontinuity, a Hot Flow anomaly (HFA) develops at contact with the shock because the shock reflected ions are accelerated into the discontinuity where they accumulate. This obviously happened at Saturn as shown in Figure \ref{chapBS-fig-CassHFA}, and the Saturnian bow shock reacted in about the same way as the terrestrial bow shock by evolving Hot Flow Anomalies. These events were observed by {\small Cassini} during its first two orbits around Saturn. The Saturnian HFA is somewhat different from those observed at Earth as it has two magnetic walls at its boundaries while the near-Earth HFAs always possessed strong shocks which confined the current and plasma in its interior. Nevertheless, the observation of this structure shows that the Saturnian bow shock behave in some respect similar to the terrestrial bow shock.
\begin{figure}[t!]
\vspace{-0.4cm}\centerline{\includegraphics[width=0.7\textwidth,clip=]{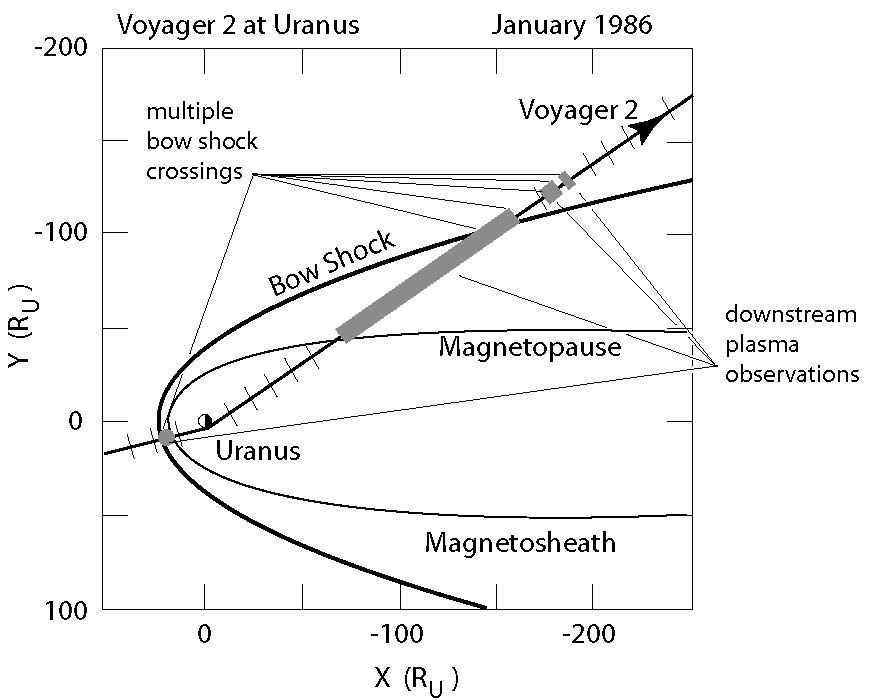}}
\caption[Saturn] 
{\footnotesize The path of Voyager 2 around Uranus in January 1986. Indicated are the multiple bow shock crossings as seen in the plasma instrument \citep{Bridge1986}. The grey bars indicate the observation of thermalized magnetosheath plasma downstream of the bow shock which is bound by the shock and gives a rough indication of their geometry. Note the flapping of the bow shock at the flank crossings at January 27-29. }\label{chapBS-fig-Uranus}
\end{figure}

\subsubsection{Uranus' bow shock}\index{shocks!Uranus' bow shock}
\noindent The planet Uranus is particular among the outer giant planets because its magnetic axis is inclined towards the ecliptic by an angle of 59$^\circ$ against its rotation axis. Rotating faster than Earth and being of four times larger diameter, the magnetosphere\index{magnetosphere!Uranus} is in a fast wobbling motion, flapping up and down with respect to the ecliptic. Moreover the magnetosphere has a strange configuration with the poles being so close to the ecliptic, which makes the cusps of the magnetosphere much more vulnerable to direct solar wind inflow and to widening than on Earth. Probably these conditions do not have very strong effects on the Uranian bow shock, however, as the conditions at the bow shock are determined mainly by the gross properties of the obstacle. The distance between the shock and the magnetospheric obstacle is very large, comparable to Jupiter and Saturn (see Table \ref{chapPBS-table2}), such that one does not expect that the behaviour of the magnetosphere directly transfers to the bow shock. The Uranus bow shock is probably controlled by the solar wind. One, thus, does not expect to encounter large differences between the bow shock of Uranus and those of Saturn and Jupiter, except for the fact that the solar wind is more dilute, colder than at the latter planets, and the magnetic field is weaker, which implies a lower plasma $\beta$ and larger Mach number. 

Uranus was passed by {\small Voyager 2} in January 1986 (see Figure \ref{chapBS-fig Uranus}). A first contact with its environment was dated on 22 January 1986 at distance 140 R$_{\rm U}$ when {\rm Voyager 2} detected a burst or 28-keV ions coming from the planet which lasted for $\sim$12 hours \citep{Krimigis1986}. Two days later, just before bow shock encounter, these fluxes increased by two orders of magnitude, briefly decreasing after shock passage and increasing to $10^3$ times the interplanetary flux level after entering the Uranus magnetosphere. The bow shock crossing itself was not spectacular in the particle fluxes, however. Instead, the bow shock crossing was identified in magnetic field \citep{Ness1986} and plasma wave observations \citep{Gurnett1986}. In the latter the shock crossing occurred 0728 spacecraft time on 24 January 1986  at planetocentric distance of 23.5 R$_{\rm U}$ as a 4 min long broadband electric wave burst that was preceded hours before by electron plasma oscillations in the 1.78 kHz channel in Langmuir waves, signalling magnetic connection and the emission of electron beams from the shock and fixing the pre-shock solar wind plasma density to a surprisingly high value of $N\sim 4\times10^4\,{\rm m}^{-3}$, three times higher than expected at the Uranian distance of 19 AU. Somewhat surprising was the low plasma wave activity in the shock foot which \cite{Moses1989} attribute to the unusually high plasma-$\beta$ during this upstream crossing comparably near to the Uranian bow shock nose. Again the strong broadband burst is an indication of the two stream instabilities acting in the shock ramp transition and overshoot which produces ion heating and plasma scale localised wave structures. Similarly to the other giant planets passed by {\small Voyager}, the wave activity in the adjacent downstream magnetosheath was very low. This first crossing has been identified as a quasi-perpendicular shock \citep{Bridge1986} in a high Mach number plasma with plasma $\beta\sim 1$.
\begin{figure}[t!]
\vspace{-0.2cm}\centerline{\includegraphics[width=0.7\textwidth,clip=]{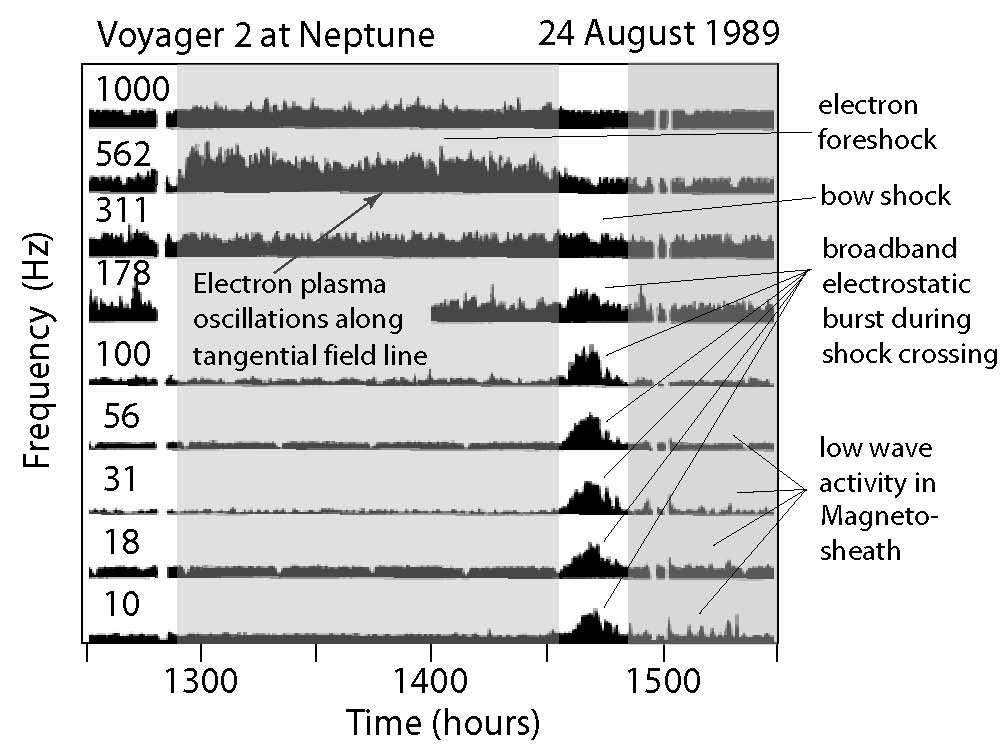}}
\caption[Saturn] 
{\footnotesize The Voyager 2 encounter of the Neptunian bow shock on 24 August 1989 as seen in the plasma wave recordings  \citep[after][]{Gurnett1989}. Shown are the electric wave field intensities in black shadowing. The intensity scale  in each of the frequency bands shown is three orders of magnitude from $10^{-6}$ mV/m to $10^{-3}$ mV/m. From 1300 Spacecraft time on until hitting the bow shock Voyager 2 was in the electron foreshock as indicated by the intense electron plasma oscillations emitted along the shock tangential magnetic field line. The shock encounter around 1520 (left white in the figure) appears as the intense broadband noise up to 311 Hz. Afterwards the spacecraft is in the surprisingly quiet Neptunian magnetosheath. }\label{chapBS-fig-Neptune}
\end{figure}

\subsubsection{Neptune's bow shock}\index{shocks!Neptune's bow shock}
\noindent The outermost planet, also a giant, is Neptune. He was the lasts to be visited by manmade spcecraft, when on 24 August 1989 {\small Voyager 2} reached it on its way out of the solar system. As suspected, Neptune was found to possess a sstrong and complex intrinsic magnetic field, a magnetosphere\index{magnetosphere!Neptune} in interaction with the solar wind, and a bow shock standing at an upstream distance of 34.9 R$_{\rm N}$ \citep{Ness1989,Belcher1989}. This bow shock was left at the tail flanks where its signature was still detectable. The sunward comression factor of the shock was $B_2/B_1\sim 2-3$ indicating a fairly strong shock. It also possessed a susceptible overshoot, a foot region and some magnetic fluctuations in front of it. Thus the shock was quasi-perpendicular. Not surprisingly, the plasma wave instrument aboard {\small Voyager 2} observed all the features that are common at a supersonic (Mach number $\sim$6) colisionless bow shock crossing: upstream plasma oscillations indicating magnetic connectivity to the shock and electron beams emitted from the shock, broadband electrostatic turbulence in the bow shock crossing and transition region, and a drop to the instrument noise level in the adjacent magnetosheath \citep{Gurnett1989}, similar to what had been seen in the other giants, Saturn and Uranus. The plasma oscillations were in the 562-Hz channel which corresponds to an upstrem plasma density of $N\sim 3.9\times10^3\,{\rm m}^{-3}$. Figure \ref{chapBS-fig-Neptune} shows the plasma wave recording during this first encounter of the Neptunian bow shock.

\section{Mars and Moon}\index{planets!Mars}\index{moons!Moon (Luna)}
\noindent Mars and Moon are the two celestial bodies in our nearer spatial environment which either have no or a very small magnetic field, no or a very thin atmosphere, which come into direct contact with the solar wind and are large enough (see Table \ref{chapPBS-table2}) to distort the solar wind flow as their diameters are larger than the gyroradius of a solar wind ion. 

\subsection{Lunar mini-bow shocks}\index{shocks!Moon's mini-bow shocks}
\noindent The typical ion solar wind ion gyroradius at Moon is of the order of $r_{ci}(1\,{\rm AU})\sim$ (400-1000) km depending on the magnetic field strength. Hence, if the Moon possessed an own magnetic field or if the passing solar wind would induce a magnetic field in the body of the Moon, this lunar field would deflect the solar wind, cause a magnetosphere around Moon and a detached bow shock standing in front of its magnetosphere. The size of the magnetosphere and stand-off distance of this bow shock would depend on the strength of the interior or induced fields. As long as nothing is known about the proper interior field, nothing can be said about the presence of such a magnetosphere and/or bow shock \citep{Sonett1967}. However, an induced field would at most be of the same order as the solar wind field depending on the lunar conductivity. 

Thus the first observation of the lunar magnetic field was expected with great curiosity. These observations were provided by the {\small Lunar Orbiter}\index{spacecraft!Lunar Orbiter} {\small Explorer 35}\index{spacecraft!Explorer 35} in 1967 with the result that the Moon does indeed disturb the solar wind generating a plasma diluted wake behind its body while the Moon does not show any indication of a magnetosphere and definitely does not possess a bow shock \citep{Colburn1967}. 

This finding is of interest as it tells that the Moon does also not possess any significant lunar magnetic field, whether intrinsic or induced. The absence of any remarkable induced field, in addition, tells something about the value of the bulk lunar resistivity which must be larger than $10^6$ Ohm\,m for an interplanetary magnetic field strength of 5 nT \citep{Sonett1967}. The Moon absorbs part of the impacting solar wind, part of it passes the Moon like a dielectric medium, and the distortions that the solar wind and its magnetic field experience are entirely due to the deflection of the frozen-in solar wind magnetic field and particles around the Moon. The Moon creates a shadow in the solar wind screening its backside from the solar wind particles which have not enough thermal spread in order to fill the gap behind the Moon. The minimum length of this shadow in lunar radii is $\sim$6 R$_{\rm L}$. This particle shadow, which appears as a lunar wake is bound to its sides by current sheets of width the order of an ion gyroradius. Their task is to cancel the magnetic field in the region behind the Moon that is void of plasma. They are dynamically active and cause distortions of the magnetic field and solar wind plasma from current drift instabilities behind the Moon, but no bow shock wave is generated in front of the Moon. 

Transverse fluctuations in the magnetic field that have been detected upstream and downstream near the Moon \citep{Ness1968,Ness1969} with frequencies in the range $0<f\lesssim 1$ Hz on magnetic field line that cross the Moon's wake have been shown to result from just these instabilities. Electrons that are accelerated in the unstable wave fields escape along the magnetic field to upstream of the Moon, where they cause ballistic instabilities \citep{Krall1969} and excite electron plasma oscillations which mistakenly can be taken as the signature of a lunar bow shock in similarity to such signatures upstream of the bow shocks of the magnetised planets.
\begin{figure}[t!]
\vspace{-0.2cm}\centerline{\includegraphics[width=1.0\textwidth,clip=]{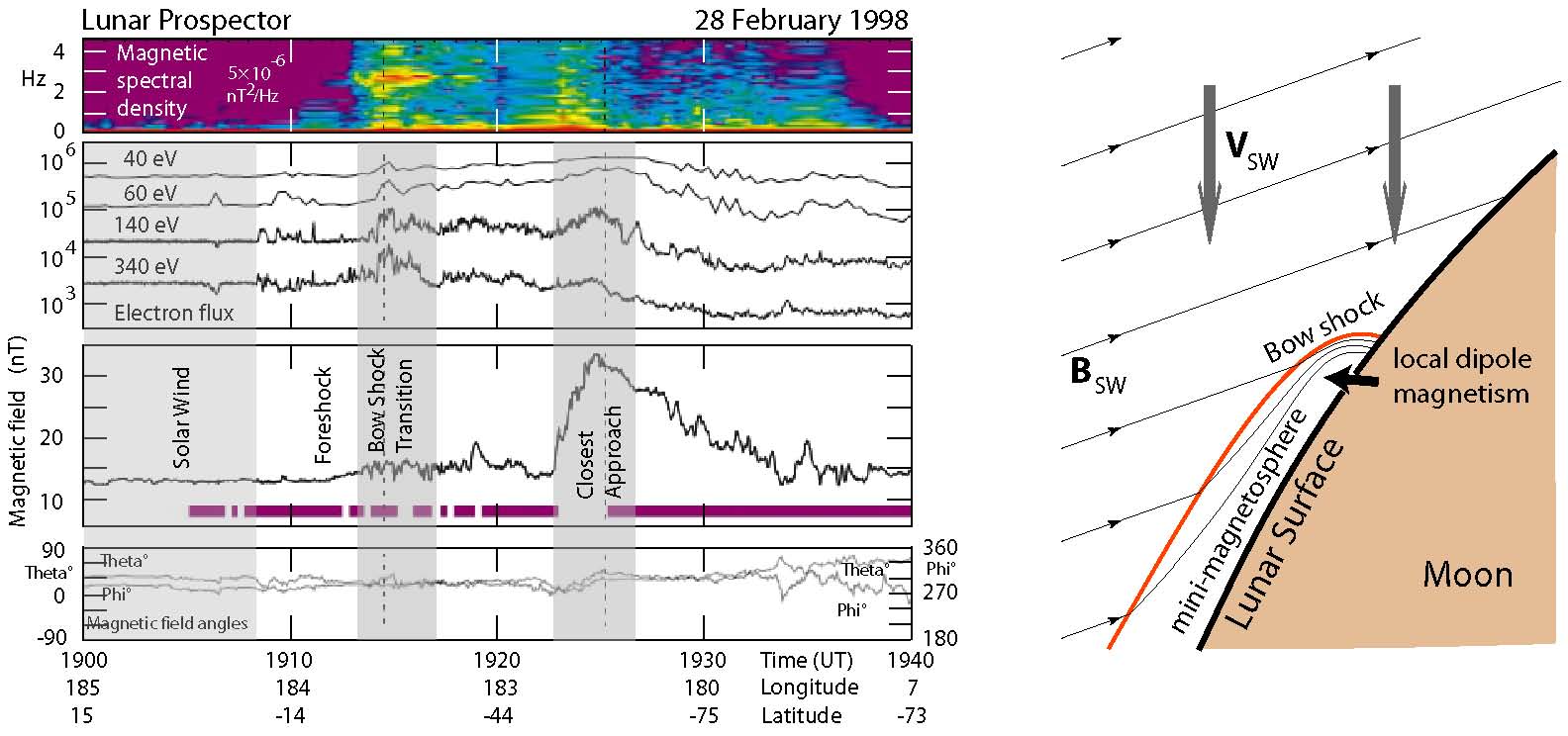}}
\caption[Saturn] 
{\footnotesize The passage across a mini-magnetosphere close to the lunar surface by the Lunar Prospector on 28 February 1998 \citep[after][]{Lin1998}. The spacecraft passes from the solar wind into a region similar to the electron foreshock with energetic electron beans emanating from the lunar direction along the magnetic field to the spacecraft. Afterwards a shock-like discontinuity is crossed which has the character of a detached bow shock which shields a strong magnetic field from a local crustal magnetism on the Moon. The low frequency wave spectrum shows the magnetic fluctuations that are related to the foreshock and bow shock. Four energy channels of electrons are shown. The two bottom panels give the magnitude and two angles of the magnetic field. The fiel reaches 30 nT. On the right the model of the mini-magnetosphere with its mini-bow shock is sketched.}\label{chapBS-fig-Moon}
\end{figure}

Apart from the absence of any large-scale intrinsic lunar magnetic field, its consequences for the internal dynamics of the Moon, and the limitations on a global lunar conductivity \citep{Sonett1975} that are set by the absence of a lunar bow shock wave, it cannot be concluded that the Moon does not have any local magnetic fields which result from remanent magnetisation of its rocky material or otherwise from highly conducting materials in which magnetic fields of solar wind strength could be induced when they become exposed to the solar wind magnetic field passing the Moon. Such localised internal magnetic fields may well cause mini-magnetospheres which are screened against the solar wind by min-bow shocks that exist only in the region where the lunar magnetisation occurs. Situations of this kind have recently been investigated by numerical simulation \citep{Harnett2000} after measurements performed during the {\small Apollo} missions suggested that such local fields up to strengths of 10$^3$ nT an extensions $>$100 km are indeed present on the Moon  \citep{Sharp1973, Dyal1974,Hood1980,Hood1981,Lin1988} and the {\small Lunar Prospector} mapped the lunar surface magnetic field and found surfac field $\lesssim $300 nT over extensions of $\lesssim$ 1200 km \citep{Lin1998}. The condition for such magnetosphere/bow shock systems to exist is related to the ion gyroradius, i.e. the extension of the magnetised region must be substantially larger than the solar wind ion gyroradius. \index{spacecraft!Lunar Prospector}

\begin{figure}[t!]
\centerline{\includegraphics[width=0.7\textwidth,clip=]{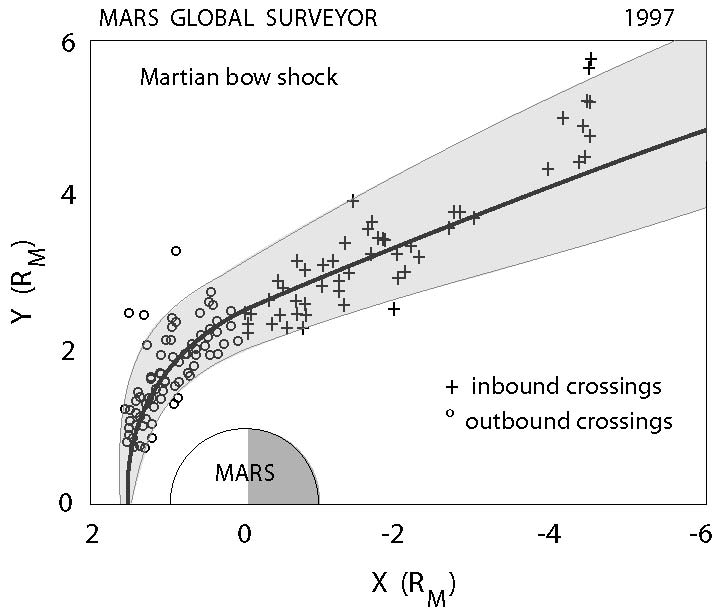}}
\caption[Saturn] 
{\footnotesize The Mars Global Surveyor passages across the Martian bow shock. Circles indicate the inbound passages which were all near the stagnation point. The outbound crossings were at the morning flank of the bow shock. Large flappenings of the shock are seen by the scatter in the data. The range of variation of the location of the Martian bow shock has been indicated by shading \citep[data taken from][]{Acuna1998}.}\label{chapBS-fig-Mars2}\vspace{-0.5cm}
\end{figure}
Figure \ref{chapBS-fig-Moon} shows a case of the {\small Lunar Prospector} observations with indication of a mini-magnetosphere and bow shock. The spacecraft moves from the solar wind into a region which resembles an electron foreshock with enahnced fluxes of electrons coming from the Moon along magnetic field lines. Then it crosses a signature that resmbles a shock transition followed by a disturbed field zone and a strong increase of the magnetic field at closest approach to Moon.\index{magnetosphere!mini-magnetosphere} 

\cite{Harnett2000} found that such mini-magnetospheres can indeed be formed under the condition that the magnetic anomaly field strength on the Moon is $>$10 nT  at an altitude of 100 km above the lunar surface, corresponding to a surface field of $\sim$300 nT whenever the solar wind density is $<4\times10^7\,{\rm m}^{-3}$. Such lunar mini-magnetospheres have a horizontal extension of the order of 100 km and are extremely vulnerable with respect to variations in solar wind pressure and magnetic field direction. They would appear and disappear, shrink and broaden and wander around depending on solar wind conditions making the lunar surface a magnetically living animal with magnetospheres and bow shocks arising and dying away in regions of high magnetisation or high conductivities.  Average observed local lunar surface magnetic field strengths at 100 km altitude are of the order of $\sim$2 nT and should not lead to the generation of mini-magnetospheres and mini-bow shocks. 

These considerations of the Moon and its localised magnetic fields can be extended also to the larger asteroids of sizes well above the solar wind ion gyroradius. If these objects are magnetised or  consist of highly conducting material, mini-magnetospheres might form around them either being supported by the internal magnetism or by the magnetic fields that can be induced in their conducting interiors. In both cases mini-bow shocks might exist around them having similar properties as the proposed lunar mini-bow shocks, deflecting the solar wind and contributing to its very localised heating.

\begin{figure}[t!]
\vspace{-0.2cm}\centerline{\includegraphics[width=1.0\textwidth,clip=]{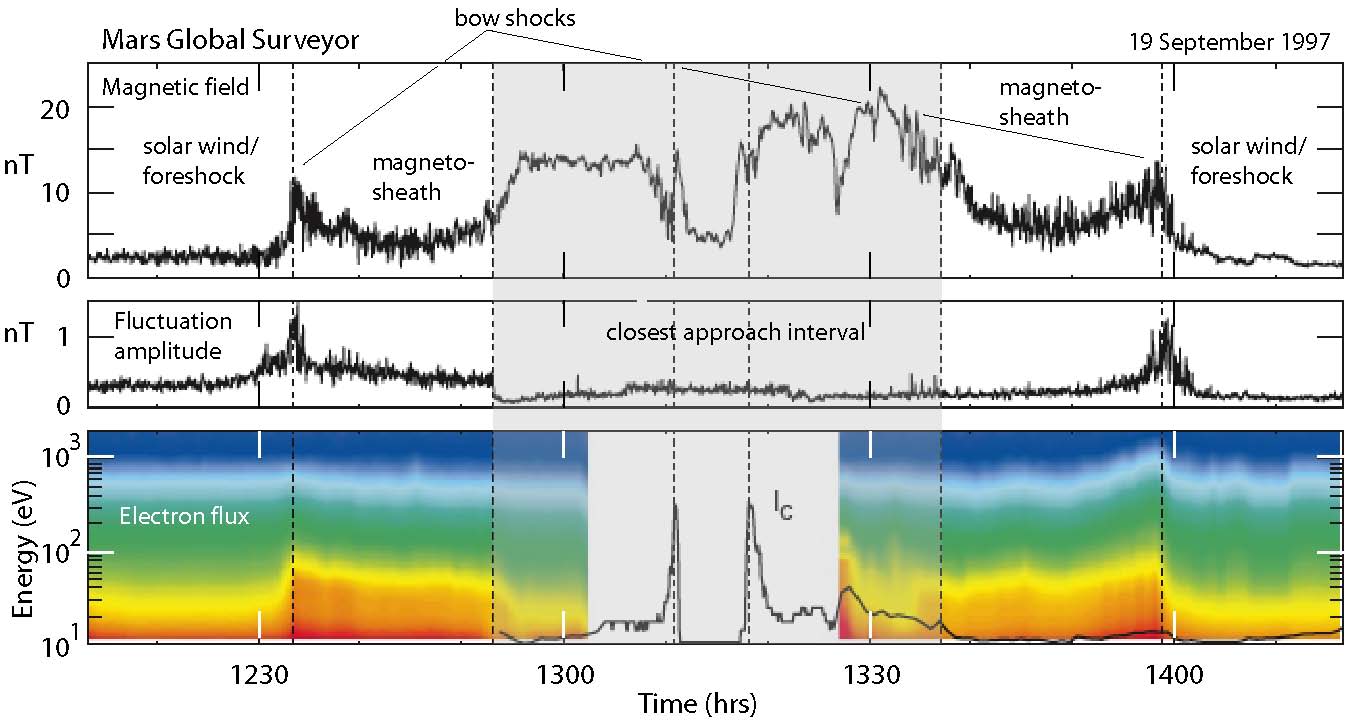}}
\caption[Saturn] 
{\footnotesize The Mars Global Surveyor passage near the red planet on 19 September 1997 as seen in the magnetic field, ultra low frequency magnetic fluctuations and electron flux measurements \citep[data taken from][]{Acuna1998}. Contact with the bow shock is preceded by increased magnetic fluctuations and increasing electron energy. Two shock crossings (inbound and outbound) are seen as steep magnetic transitions, peaked fluctuation levels and high electron thermal spreads, followed by a broad magnetosheath. The obstacle here is not the magnetic field but the Martian atmosphere.  The black line in the electron spectrogram is the Langmuir probe current which is a measure of electron densities and temperatures close to the planet. }\label{chapBS-fig-Mars1}\vspace{-0.5cm}
\end{figure}

\subsection{Mars - a pile-up  induced bow shock}\index{shocks!Mars' induced bow shock}
\noindent It has been believed for long that Mars possesses an internal magnetic field though weak but strong enough to form a magnetosphere around the Red Planet that has been named after the Roman {\it Deus Belli}. Mars was always the beloved darling of man onto that he has dreamt away to build his cities and bunkers under the Martian crust after having escaped from Earth. Previous spacecraft like {\small Phobos} and {\small Mars Pathfinder} \index{spacecraft!Mars Pathfinder}\index{spacecraft!Phobos} could not definitely exclude the existence of a Martian magnetic dipole field. The proof that Mars has no global internal field was left to the {\small Mars Global Surveyor} \citep{Acuna1998,Acuna2001}. The {\small Phobos} spacecraft had detected substantial ultra low frequency wave activity in the magnetic field during the Mars approach \citep{Russell1990} that was attributed to a Martian bow shock wave respectively the existence of a Martian global field. 

The {\small Mars Global Surveyor} \index{spacecraft!Mars Global Surveyor} set an upper limit of $B<0.5$ nT \citep{Acuna2001} on a global surface field on Mars. Two the great surprise, when mapping the magnetic surface field of Mars, it found that Mars possesses several cratereous regions of fairly strong local remanent magnetic fields of magnitude $B\lesssim200$ nT, indicating that in early times Mars was magnetised, had a working dynamo and left some signatures of these times in his crustal minerals. Thus, had Mars no atmosphere it would today behave very similar to the Moon, lacking a global magnetosphere\index{magnetosphere!pile-up induced magnetosheath} and a global bow shock but instead being covered with regions of mini-magnetospheres and min-bow shocks which reacted highly dynamically to the fluctuations in solar wind pressure and magnetic field.  

However, Mars has an own thin atmosphere, the existence of that was confirmed. Similar to Earth's atmosphere, the Martian atmosphere becomes ionised by the solar ultraviolet radiation and forms an ionosphere on the illuminated side of the planet which prevents the solar wind from impacting the Martian surface. Mars, in this sense, behaves similar to Venus with the main difference that its atmosphere and hence its ionosphere as well are thin. Nevertheless pressure balance between the solar wind and the ionosphere creates and ionopause and a blunt detached bow shock that stands in front of the ionopause at a planetocentric stand-off distance of $\sim$1.5 R$_{\rm M}$. 

The {\small Mars Global Surveyor} spacecraft crossed the Martian bow shock many times on its outbound and inbound passes before settling on a circular mapping orbit around Mars at altitude of $\sim400$ km. Figure \ref{chapBS-fig-Mars2} shows a collection of the identified locations of the bow shock in the projection onto the $(x,y)$-ecliptic plane \citep[data taken from][]{Acuna1998}. The inbound crossings are exceptionally at the flankside bow shock and are at distances farther away from the planet than the outbound crossings which were located on the morning side of the planet. It is interesting to see that the shape of the Martian bow shock is highly non-stationary, obviously flapping around an average position with quite large amplitude. Moreover, the bow shock seem to be more blunt on the dayside than other planetary bow shocks and might inflate even more at larger tailward distances. There are different reasons for this flapping. It may be caused by fluctuations in the solar wind pressure as well as by fluctuations in the Martian ionospheric ionisation that is affected by the atmospheric conditions on Mars. Part of it may also be affected by the average local crustal magnetic field strength over one Martian rotation which contributes to the resistance of Mars to the solar wind, but the main reasons can probably be seen in the interplay between the atmospheric and solar wind conditions.

Apart from this global shape of the Martian bow shock, the microscopic signature is physically of more interest. Figure \ref{chapBS-fig-Mars1} shows some of the {\small Mars Global Surveyor} measurements during two of the first crossings of the Martian bow shock. The top two panels give the magnetic recordings magnetic field strength and rms fluctuation amplitude versus time on 19 September 1997. The bottom panel gives the colour coded spectrogram of the electron energy fluxes (in electrons/s cm$^2$ sr eV) in the energy range from 10 eV to 2 keV. Highest fluxes are shown in red, lowest fluxes in blue. The absolute scale of the fluxes is not given here. The shaded region in the figure is the inner part of the spacecraft at closer approach to the planet which does not interest us in this connection except for the strong increase in the magnetic field before reaching the ionopause. 

The spacecraft crosses the shock on its inbound path at about 1235 Spacecraft Time. Prior to this crossing it is in the solar wind detecting enhanced fluctuation levels in the low frequencies with increasing level the closer it comes to the shock. There is a well expressed shock foot region with high and nearly constant fluctuation amplitudes from 1230-1234 Spacecraft Time. The shock crossing exhibits a steep ramp and  overshoot, both of them are also seen in the peaked fluctuation levels. These are followed by a broad turbulent magnetosheath region with enhanced but lower levels until spacecraft entrance into the increasing magnetic field which is a magnetic pile up boundary standing in front of the ionopause at the start of the shaded area. This pile up is created by the draping and compression of the interplanetary magnetic field around the highly conducting ionosphere that inductively screens the planet from the interplanetary field. It is this magnetic pile-up that acts as the obstacle for the solar wind and forces the detached bow shock to develop. Would it not exist, would the bow shock probably be attached to the atmosphere due to the unbroken penetration of the solar wind into the atmosphere. \index{boundaries!magnetic pile up}

The electron energy fluxes also show an increase in energy with entrance into the bow shock foot region, corresponding to electron heating, and a broad intense burst of electrons related to the shock crossing and showing its signature up to 1 keV. Clearly the shock is a region of plasma heating also in this case. However, in the lowest energy channels the electron fluxes have become very strong indicating that more electrons participate than are present in the solar wind, which implies that the plasma is compressed. Indeed, the compression in the magnetic field from the $\sim$2 nT solar wind field to the $\sim$(5-6) nT shocked magnetosheath field indicates a moderately strong shock compression factor of about 2-3. Note also the decrease in electron flux and cooling at the magnetic pile up boundary.

The outbound shock crossing at 1358 is similar but with a much longer and better developed foot region, compression of roughly a factor 3, higher overshoot exceeding 10 nT and a broad downstream magnetosheath region which, however, is less turbulent than the inbound magnetosheath as seen from the rms amplitude. The fluctuations concentrate solely on the shock transition and parallel the stronger electron heating which now extends to electrons $>$1 keV. We should note that the first inbound crossing was a tail bow shock crossing while the outbound crossing is a dayside crossing that is closer to the planet in the stronger compressed magnetic field and plasma. In both cases, however, the shocks are still similar to the known shocks of the other planets, both being quasi-perpendicular. This must be so in the case of Mars because the Martian ionosphere almost screens the solar wind magnetic field from the planet. This Mars has in common with his sister planet Venus.
\begin{figure}[t!]
\vspace{-0.2cm}\centerline{\includegraphics[width=1.0\textwidth,clip=]{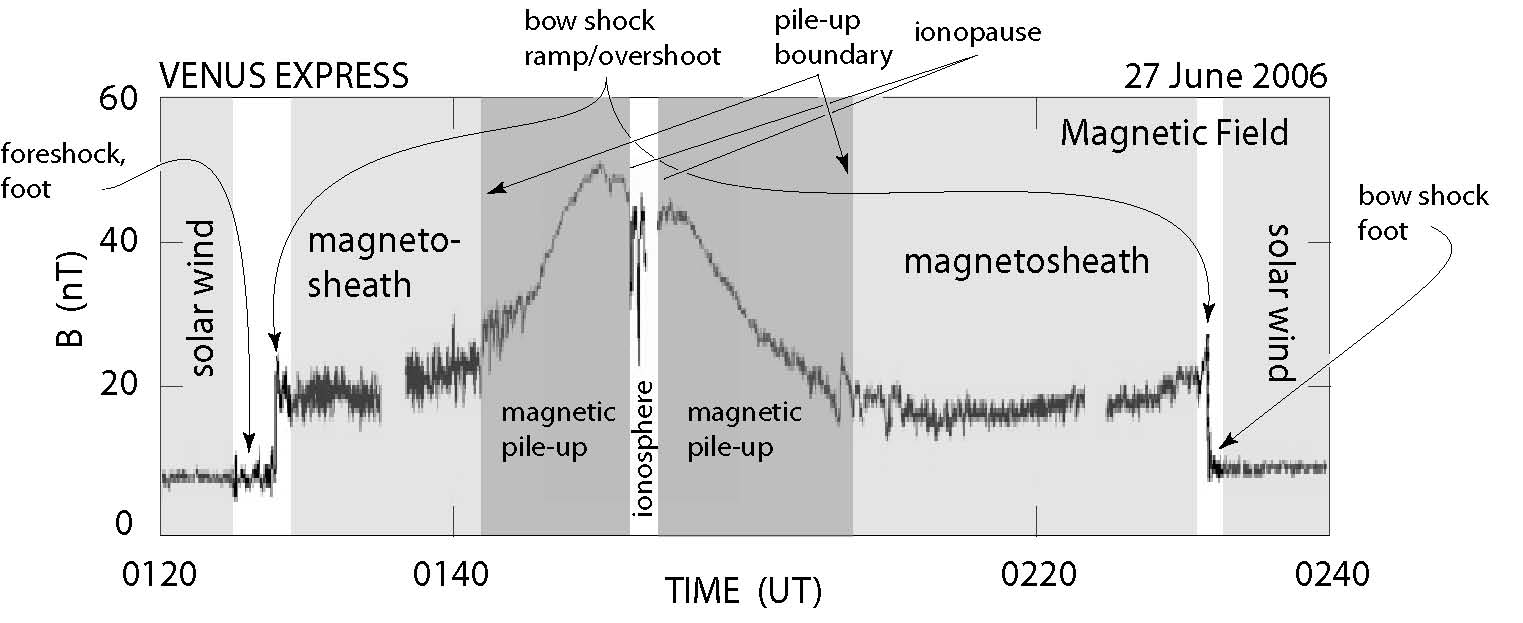}}
\caption[Saturn] 
{\footnotesize Recent magnetic field recordings during a Venus Express passage of the Venusian bow shock \citep[data taken from][]{Zhang2008}, showing the various different regions and magnetic boundaries in the environment of Venus. During inbound the spacecraft detected a weak foreshock exhibiting magnetic fluctuations, a short foot region, a quasi-perpendicular bow shock ramp and overshoot, entered a disturbed magnetosheath and passed the magnetic pile-up boundary until reaching the ionopause. On the outbound path no foreshock was detected except for a short shock foot.}\label{chapBS-fig-Venus2}\vspace{-0.5cm}
\end{figure}

\section{Venus and the Comets}\index{planets!Venus}
\noindent In contrast to the other planets and the Moon, Venus and the comets have in common that they are non-magnetic and both have dense and extended atmospheres even though their atmospheres are of completely different origin. Thus the interaction of Venus and the comets with the solar wind is basically different from that of the other planets. One, however, expects that Venus gives rise to the formation of a strong shock wave that stands in front of the planetary atmosphere either as a detached or atmospherically attached wave. The reason is that Venus is large enough to deflect the super-magnetosonic supercritical solar wind around its body and dense extended atmosphere while the velocity of compressional waves that are needed to maintain the deflection of the solar wind is too small to compete with the solar wind flow. Whether comets behave the same way like Venus or not, depends, however, on their particular properties. Before coming to discuss some of the observations of the cometary interaction with the solar wind in the inner heliosphere we turn to Venus and what has been learned about its interaction with the solar wind from the various observations {\it in situ} its environment.

\subsection{The Venusian bow shock}\index{shocks!Venus' bow shock}
\noindent A venus-like interaction is defined as an interaction that is associated with a planet that has no intrinsic magnetic field but a substantial atmosphere. Indeed, the upper limit for the intrinsic magnetic field of Venus (set by the {\small Pioneer Venus Orbiter after initial measurements by the Venera satellites) is $\sim 10^{-5}$ times\index{spacecraft!Pioneer Venus Orbiter} that of Earth\index{spacecraft!Venera}. Venus also does not have any remanent crustal magnetic field left over from an early active magnetic dynamo period, because the temperatures in Venus' crust are expected to lie above the Curie point. These high temperatures imply complete destruction of any remanent magnetism. Under these conditions the solar wind can reach directly to the top of the ionosphere, assuming that an ionosphere exists and has not to be created by the impact of the solar wind. This has been the case already for Mars as we have discussed above. Venus, however, has a much denser atmosphere than Mars and, in addition, is much closer to the Sun (cf. Table \ref{chapPBS-table1}). Being at about half the distance of Mars the solar radiation\index{radiation!solar at Mars} is four times more intense than at Mars and 1.6 times more intense than at Earth. The dense Venus atmosphere absorbs this optical and extreme ultraviolet radiation and produces a highly conducting ionosphere at altitudes below 300-800 km with maximum ionisation between 125-170 km altitude. In the absence of the ionosphere Venus could possess an induced magnetic field if its planetary conductivity would be high enough. However, this is prevented by the presence of the ionosphere that screens the lower atmospheric layer and the planet from the solar wind and interplanetary field. Any induced field will be mediated through the ionosphere and thus be of little effect. The interaction between the solar wind and Venus is thus directly with the Venusian ionosphere \citep[for reviews see, e.g.,][]{Luhmann1986,Luhmann1991} which forms the obstacle to the solar wind. The surface of pressure balance between solar wind ram pressure and ionospheric thermal plasma is the ionopause. Farther out than the ionopause the solar wind magnetic field piles up, a blunt bow shock \citep[see, e.g.,][]{Wolfe1979,Scarf1979c} is generated at a distance upstream leaving a magnetosheath like region between the shock and the ionopause. At Venus the stand-off distance of the bow shock is found at a planetocentric distance of $\sim$1.5 R$_{\rm V}$ at solar maximum and depends on solar activity as this regulates the ionospheric electron content. In the magnetosheath the solar wind is deflected around the ionopause in a way very similar to the deflection around a magnetopause. Interestingly enough, small-scale interplanetary magnetic flux tubes can sometimes break through the ionopause being found deep in the atmosphere where they form so-called flux ropes which remain to be connected to the solar wind magnetic field and are convected around the planet.
\begin{figure}[t!]
\vspace{-0.2cm}\centerline{\includegraphics[width=1.0\textwidth,clip=]{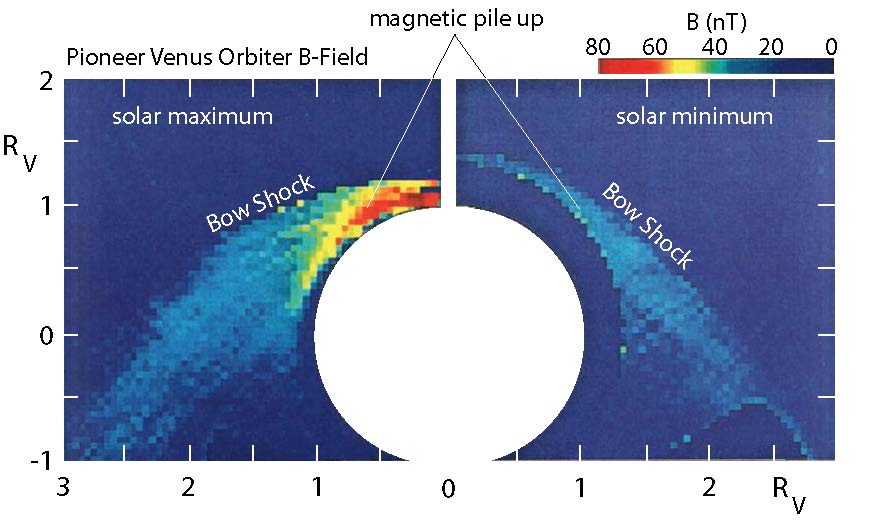}}
\caption[Saturn] 
{\footnotesize The Pioneer Venus Orbiter mangetic field measurements near Venus during solar maximum and solar minimum \citep[data taken from][]{Kallio1998}, showing the control of the magnetic pile up region and bow shock position with solar activity. At high solar activity the pile up is much stronger and the bow shock stand-off distance smaller than at low solar activity. }\label{chapBS-fig-Venus1}\vspace{-0.5cm}
\end{figure}

Being at about half the heliocentric distance of Mars the solar wind density at Venus is four times higher, and the solar wind dynamic pressure is twice that at Earth's orbit. The magnetosonic Mach number at Venus is ${\cal M}_{ms}\approx5.1$ while the solar wind $\beta$ is slightly less than at 1 AU. In the average the solar wind at Venus is super-critical but not of vastly high Mach number. On the other hand, the spiral solar wind magnetic field is in the average more inclined at Venus than at the outer planets such that one expects that its prospective bow shock possesses a fairly large quasi-parallel area, depending on the strength of the magnetic deflection in the magnetosheath. Such an interaction does not produce a magnetotail like on Earth because there is no planetary magnetic field which could be stretched out into space. An important point is that the magnetic field shows a strong shock of compression ratio $\sim$3 in both cases with sharp overshoots. Such shocks should reflect particles upstream along the tangential field and have foot regions. The latter are indeed seen in these crossing, and the inbound crossing indicates also the presence of a forshock as can be judged from the relatively large amplitude magnetic fluctuations detected $\sim$ 4-5 minutes before shock entrance.

Backstreaming ions from the Venus bow shock along the magnetic field have been reported in earlier \citep{Frank1991,Williams1991} passages of other spacecraft, in particular from the {\small Galileo} \index{spacecraft!Galileo} spacecraft on 10 February 1990. Such ions have been observed in the  keV energy range \citep{Moore1989} and might be the seed population for the more energetic ions in the 120-280 keV energy range seen by {\small Galileo} \citep{Williams1991} after having been accelerated at the Venusian bow shock. Whether this is possible, is not known as the energetic ions have gyroradii larger than the curvature radius of the small Venus bow shock. It is s¬thus more probable that these ions have been accelerated in a different way, either directly already in the Venusian ionosphere, or by Venus lightning, or by bluncing arond in the piled-up magnetic field of Venus, in which case the Venusian bow shock would not be responsible for their existence and would produce only the low energy keV ions.

 On the other hand, the search for pick-up ions leaking out from the Venus atmosphere and ionosphere into the solar wind gave only an upper limit of $N_{pui}\lesssim 10^3\,{\rm m}^{-3}$ for the pick-up ion density in front of the bow shock. This is still consistent with the scale height of the Venus atmosphere. The emasurements could be performed only outside the magnetosheath. Earlier observations \citep{Nagy1981} by {\small Pioneer Venus} in the magnetosheath \index{spacecraft!Pioneer Venus} had given indication for the presence of a substantial number of pick-up ions behind the bow shock which are accelerated in the convection field as well as the strongly  fluctuating electromagnetic field in the magnetosheath. 

The electron observations by Galileo, on the other hand, confirmed that the Venus bow shock is quite strong exhibiting density increases of a factor $\sim 3$, the same as concluded from the magnetic observations. However, only little electron heating $\lesssim 20\%$ was observed in the {\small Galileo} shock crossings, which contrasts with the bow shocks of magnetised planets and is also not in agreement with the observations at Mars dealt with above where substantial electron heating by a factor of 2-3 has been found. 

Venus has been passed since the 1960s by a substantial number of spacecraft. The most recent mission to Venus was {\small Venus Express}. Figure \ref{chapBS-fig-Venus2} shows an example of the high resolution magnetic field recordings by {\small Venus Express}\index{spacecraft!Venus Express} on 27 June 2006 with two bow shock crossings \citep{Zhang2008}. The first crossing being preceded by a short encounter with the Venusian foreshock, followed by the passage of the shock foot and the ramp and overshoot of the bow shock. Afterwards the spacecraft was in the magnetosheath until passing the magnetic pile-up boundary, pile-up region and ionopause followed by an outbound repetition of the same sequence. In both cases the shocks were quasi-perpendicular as indicated by the sharp shock fronts, the overshoot and the foot regions.  This passage is very similar to that seen above at Mars showing the close relationship between the bow shocks on these two otherwise so different planets. Similar observations were made by {\small Galileo} in February 1990 during the spacecraft encounter with Venus \citep{Kivelson1991}. {\small Galileo} crossed the shock and mapped its width to be shorter along the interplanetary field than transverse to it, confirming the near perpendicular shock structure. Moreover, small Galileo} also found large amplitude ultra-low frequency magnetic waves in front of the bow shock the amplitudes of which peaked in the magnetosheath and were attributed to the presence of pick-up ions. it was also speculated that some of the structures found might be intermediate shock waves which could, however, not been proved definitely and thus is highly speculative.

A compilation of long-term magnetic observations of the Venusian bow shock by the {\small Pioneer Venus Orbiter} is shown in colour code in Figure \ref{chapBS-fig-Venus1}. This figure shows the spatial extension of the region between the bow shock and the ionopause of Venus for solar maximum and solar minimum in the local amplitude of the magnetic field reflecting the control of the Venusian bow shock, external magnetic field and ionosphere by the solar activity. At solar minimum the bow shock is found at planetocentric distance substantially farther out than during solar maximum. On the other hand, during solar maximum the magnetosheath and magnetic pile-up\index{shocks!pile-up at Venus} regions are stronger compressed and more variable.  The reasons for these differences can be found in the fact that during solar maximum the solar wind fluctuates more strongly and has higher  Mach number and dynamic pressure causing a stronger compression of the ionopause and a stronger pile-up of the magnetic field than during solar minimum when the solar wind is relatively quiet. Moreover, during solar minimum the sun is hotter which causes a stronger ionisation of the Venusian atmosphere and thus an increased plasma density and thermal pressure in the ionosphere which lets the ionosphere expand farther out of the planet.

We finally turn to the observation of high frequency waves during the {\small Galileo} Venus bow shock encounters \citep{Gurnett1991,Hospodarsky1994}. The plasma wave instrument on the {\small Galileo} spacecraft detected the shock in both the electric and magnetic instruments as broadband impulsive noise similar to what is observed in other shock crossings, followed by enhanced plasma wave turbulence in the adjacent magnetosheath. Close upstream to the shock an intense electromagnetic emission band between 5-50 Hz was detected. the nature of this band is consistent with the assumption of electromagnetic whistler or ion cyclotron waves. Since no suprathermal particle component could be identified from the plasma instruments during this time these waves are probably caused at the shock from where they propagate to spacecraft position in the solar wind. The geometry of how this can happen is not clear, however, as the solar wind is super-Alfv\'enic. \index{shocks!pile-up induced}\index{turbulence!magnetosheath}

\begin{figure}[t!]
\vspace{-0.2cm}\centerline{\includegraphics[width=1.0\textwidth,clip=]{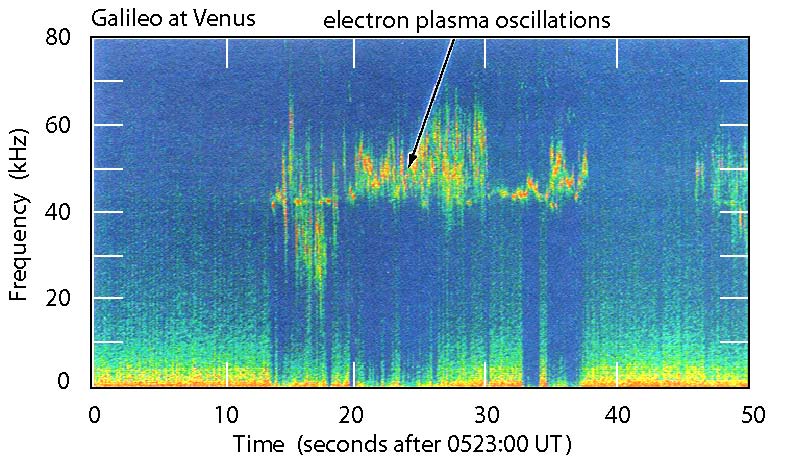}}
\caption[Saturn] 
{\footnotesize Galileo observations of electron plasma waves in the Langmuir mode as are excited by electron beams emitted from the Venusian bow shock into the electron foreshock. intensification of the waves occurs when the spacecraft crosses the tangential magnetic field line \citep[after][]{Gurnett1991}. Note the high time variability which indicates wave localisation. The disappearance of the low frequency ion acoustic waves in response to the Langmuir waves is also of interest.}\label{chapBS-fig-Hosp}\vspace{-0.5cm}
\end{figure}

Electron plasma waves at the higher frequencies of 10-50 kHz in the Langmuir wave band have been observed upstream of the shock in complete analogy to other bow shocks.  An example from the {\small Galileo} observations is shown in Figure \ref{chapBS-fig-Hosp} exhibiting a very close similarity of these oscillations to those observed in front of Earth's bow shock and also at the other magnetised planets. These waves had been seen before by {\small Pioneer Venus} in May 1979 \citep{Scarf1980}. \index{Scarf, Frederic L.} They consist of two distinct types, the Langmuir waves which map the plasma frequency at roughly 45 kHz and intensify in the presence of a fast electron beam, and a sporadic upstream downshifted emission which signals the presence of suprathermal electrons and disappears abruptly when the electron foreshock boundary is crossed into the undisturbed solar wind. The electrons come from the bow shock where they must have been accelerated. Our knowledge of the reflection and acceleration mechanism tells us that the Venusian bow shock contains highly localised electric field structures in order to produce such intense suprathermal electrons. It thus behaves very similarly to the bow shocks of the magnetised planets. \cite{Hospodarsky1994} investigated these oscillations in more detail finding the expected close similarity to the plasma oscillations and electron plasma waves in the Earth's electron foreshock. These waves are strongly localised and form structures of Debye length scales, in this case the order of $\sim 60-70$ m,  which are generated by nonlinear interaction between the suprathermal shock-accelerated electrons and the plasma wave component. 

We may thus conclude, that under normal solar wind conditions  at the Venus orbit (i.e. high speed solar wind, nominal plasma densities and magnetic field strengths) the bow shock behaves similar to the supercritical bow shocks of magnetised planets, i.e. the upstream flow does not care of what kind the obstacle is as long as the obstacle consists of a conducting plasma layer which in this case (and the case of Mars) is the ionosphere. This layer forces the magnetic field to pile-up upstream of it thereby inducing a magnetic obstacle which, however, is itself of purely upstream nature, while the ionosphere becomes a screening current layer. This obstacle deflects the solar wind  and generates a strong supercritical bow shock wave that is detached from the Venusian ionosphere. 

However, occasionally the normal conditions at Venus become violated when slow dense plasma streams encounter the planet. Such cases have recently been reported from {\small Venus Express}, when the spacecraft crossed the Venus bow shock several times under subcritical conditions with Mach number ${\cal M}<1.5$ at shock normal angles $\thetabn>70^\circ$ when no ion reflection takes place \citep{Balikhin2007}. Such cases have not been described in this book as in the heliosphere they are found very rarely and are expected to occur only in its innermost part and under extreme dense slow solar wind conditions. The upstream solar wind at the times under question was indeed unusually dense. Nonetheless, the conditions at the bow shock were still collisionless such that the shock was probably not maintained by resistive dissipation. The magnetic shock profile was of a kind similar to a Korteweg-deVries shock with either upstream or downstream attached spatially damped oscillations. Since $\beta$ was small, such a shock should be maintained ballistically with the complete cold solar wind ion beam gyrating in bulk with negligible thermal spread in the shocked magnetic field. We do not go further into this kind of observation as it represents a rare, though interesting, case of subcritical shocks\index{shocks!subcritical}  that will be realised only in very dense and comparably slow plasma streams which happen to exist in proximity to the sun. Such shocks might exist more frequently in the solar corona or outside our cosmic environment near exoplanets. Also Mercury may sometimes develop this kind of a bow shock when immersed into a dense and slow solar wind stream as generation of such subcritical shocks does not depend on the existence of a neutral atmosphere but only requires the presence of an obstacle like Venus' ionosphere and induced magnetic pile-up or Mercury's magnetosphere. \index{particles!neutral}

\subsection{Cometary bow shocks}\index{shocks!cometary bow shocks}
\noindent The first claim of a possible crossing of a cometary bow shock by the {\small Pioneer Venus} spacecraft goes back to \cite{Russell1984}. Comets need not necessarily have bow shocks. Comets consist of a relatively small icy nucleus of the size of an asteroid. This nucleus is covered by a comparably thin mantle of mostly carbon atoms mixed with dust. Where this nucleus is damaged it evaporates and outgasses attributing an extended gaseous cloud to the comet. However, far away from the sun comets outgas very weakly if at all. There is no or very little photoionisation. Ionisation is mainly due to charge exchange. Thus the mass added to the solar wind by the comet is negligible. The solar wind that flows around the comet might well induce a magnetic field in the cometary body or in its highly conducting crust layer, but because of the extremely small size of the nucleus the distortion of the solar wind at those distances is miniscule and might not even exist if the particle gyroradii exceed its diameter. However, as the comet approaches the sun, solar radiation intensity increases and heats  up the illuminated part of the comet's outer shells such that the shell starts melting in several places, the ice liquefies and ultimately evaporates, causing cometary outgassing at a susceptible rate until the comet is surrounded by a huge gas cloud many times larger than the nucleus. Part of this gas envelope becomes photoionised, charge exchange between solar wind protons and the gas atoms or molecules removes gas particles from it and adds them as pick-up ions to the solar wind. Both processes lead to mass and momentum exchange between the solar wind and the gas envelope of the comet until the rate of this exchange become large enough to decelerate and deflect the solar wind in the region where the interaction takes place. The processes involved have been investigated in detail by  \cite{Biermann1967}, \cite{Cloutier1969}, \cite{Wallis1973a,Wallis1973b} and others, and it has been shown that at large addition rates of mass to the solar wind a cometary bow shock must evolve in order to maintain the deflection of the solar wind in the interaction with the cometary cloud. The reason is again very simply that the slow compressional waves in the cometary environment are insufficient to deflect the super-magnetosonic solar wind flow around the comet. Hence, a shock must be generated in front of the comet in order to break the flow and divert it around the cometary obstacle.

\subsubsection{Comet Giacobini-Zinner}\index{comets!Giacobini-Zinner}
\noindent Investigating the nature of comets and their interaction with the solar wind has for long time been one of the major fields of interest ion space and planetary physics because of many reasons not the least one the proposal that possibly comets have been responsible of transporting germs in the solar system between the planets and contribute to the spread of life. The first mission to pass by a comet was the {\small International Cometary Explorer ICE}. In 1985 it approached the comet Giacobini-Zinner passing it at a distance of $\sim$15000 km.\index{spacecraft!International Cometary Explorer ICE} 
\begin{figure}[t!]
\vspace{-0.2cm}\centerline{\includegraphics[width=0.8\textwidth,clip=]{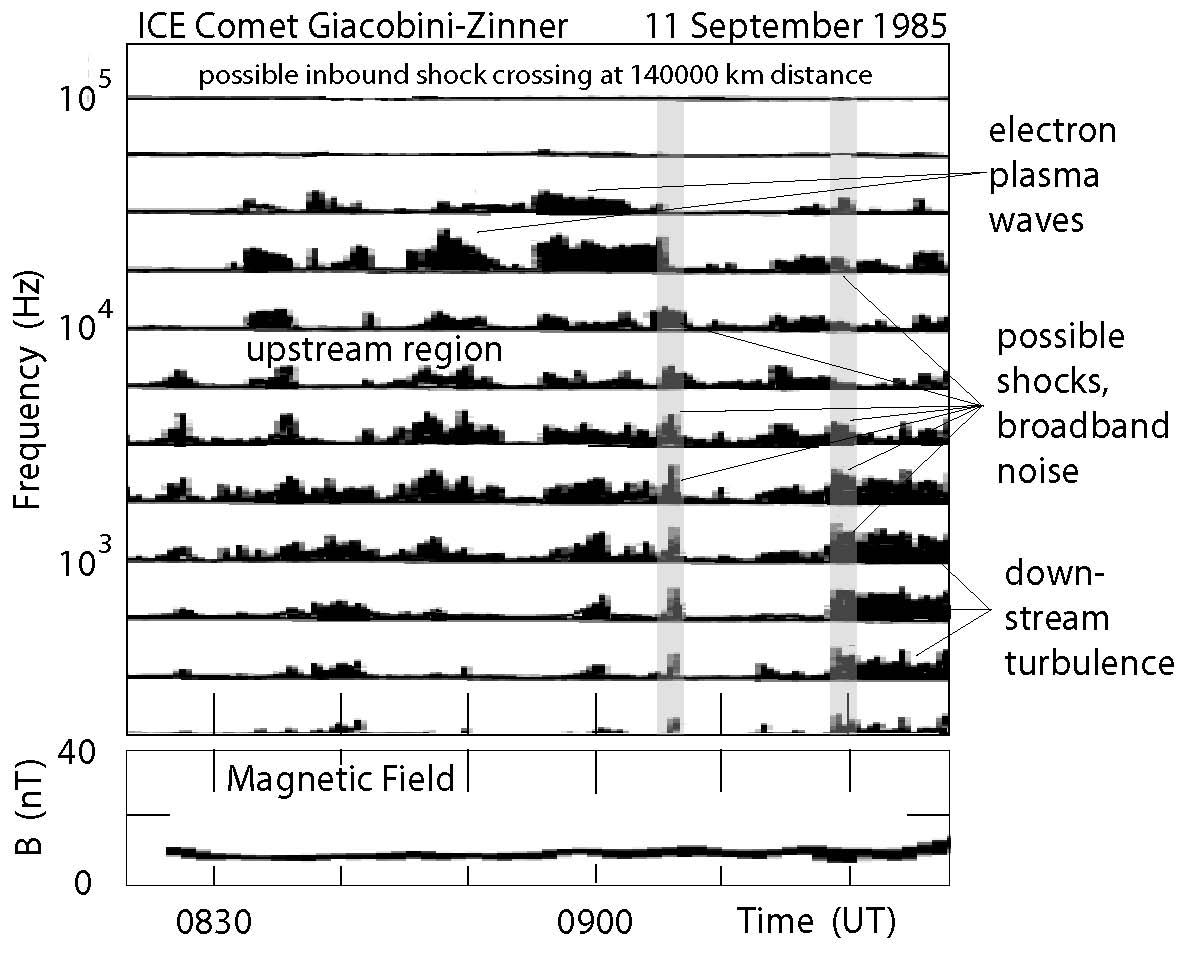}}
\caption[Saturn] 
{\footnotesize The ICE approach to comet Giacobini-Zinner on 11 September 1985 as seen in the plasma waves \citep[after][]{Scarf1986}. Before the onset of the broadband noise on the inbound passage which is typical for a shock crossing, the also typical electron plasma waves upstream of a shock were detected. Below these broader low frequency waves are seen. After the broadband impulsive shock noise the spacecraft entered the downstream magnetosheath region with intense low frquency turbulence. However, these wave observations are still inconclusive of the presence of a shock as no shock signatures were seen in the magnetic field \citep[taken from][]{Smith1986} and in the other instruments. The gradual pile-up in the magnetic field starts at about 0940 UT after the observation of the broadband noise and electron plasma waves and reaches up to 20 nT on the inbound pass. }\label{chapBS-fig-GZ1}\vspace{-0.5cm}
\end{figure}

Apart from its optical appearance on the sky, {\it in situ} information from the comet was obtained first by the {\small ICE} plasma wave instrument \citep{Scarf1986} which detected bursts of strong ion acoustic waves almost continuously at times when {\small ICE} was closer than $4\times10^6$ km away from the Giacobini-Zinner nucleus. It also observed oscillations at the electron plasma frequency and whistler waves, all being believed to be related to the presence of heavy cometary pick-up ions as well as hot electrons within this region surrounding the comet. In particular the presence of electron plasma waves could suggest  magnetic connection to the magnetic field penetrating the cometary coma and allowing suprathermal electrons to escape from it to reach the spacecraft. However, from magnetic field observations there was no indication for such a connection. Hence the plasma waves were also attributed to the presence of pick-up ions which might indeed be the case as these should locally excite lower hybrid waves which are known to accelerate electrons into beams along the magnetic field. This whole region was indeed filled with energetic cometary ions (O$^+$ and OH$^+$) of energy exceeding 30 keV \citep{Hynds1986,Ipavich1986}. Closer to comet Giacobini-Zinner within a few 100000 km distance from the nucleus {\small ICE} measured increased wave levels and broadband signals. In fact, the signal intensity was reported to increase by more than nine orders of magnitude \citep{Scarf1986}, causing saturation of the instrument in some of the channels. At the boundaries of this inner region impulsive broadband noise was measured, preceded upstream with enhanced electron plasma waves {see Figure \ref{chapBS-fig-GZ1}) which reminded strongly of shock crossings that are accompanied by electron heating. The electron plasma oscillations indicate magnetic connectivity and generation of electron beams, which is typical for bow shocks. However, only for the second outbound crossing on this day at 1219 UT the magnetic signature of a possible shock could be identified \citep{Smith1986}, making this crossing more conclusive of a bow shock pass even though \cite{Smith1986} prefer to speak of a bow wave\index{shocks!cometary bow wave} also in this case. Because of the perpendicular direction of the interplanetary field at the time of passage one would have expected a quasi-perpendicular shock in this case such that the observed lack of a steep shock signature in the magnetic field can hardly be attributed to the passage of a parallel shock. Moreover, the plasma instrument \citep{Bame1986} did not show the expected signature of  conventional shock transition in neither of the inbound or outbound crossings attempting the authors to speak of a slow transition region instead of a shock. \cite{Bame1986} speak of plasma signatures distinctly different from a conventional magnetosonic shock transition. On the other hand a broad magnetic pile-up region was passed exhibiting large-amplitude nonlinear magnetic fluctuations that resemble the upstream pulsation of quasi-parallel shocks with steep fronts and attached high frequency oscillations. On the inner edge of the pile-up region the magnetic field showed signatures of an ionopause thus resembling the crossings of the Venusian shock-pile-up region. Hence, the cometary transition region is by far structured more complicated than the transition regions of planets. 

Thus the ICE observations of comet Giacobini-Zinner were inconclusive about the possible existence of a bow shock at a comet in the inner heliosphere. A conclusion was expected from observation of comet Halley. 
\begin{figure}[t!]
\vspace{-0.2cm}\centerline{\includegraphics[width=0.95\textwidth,clip=]{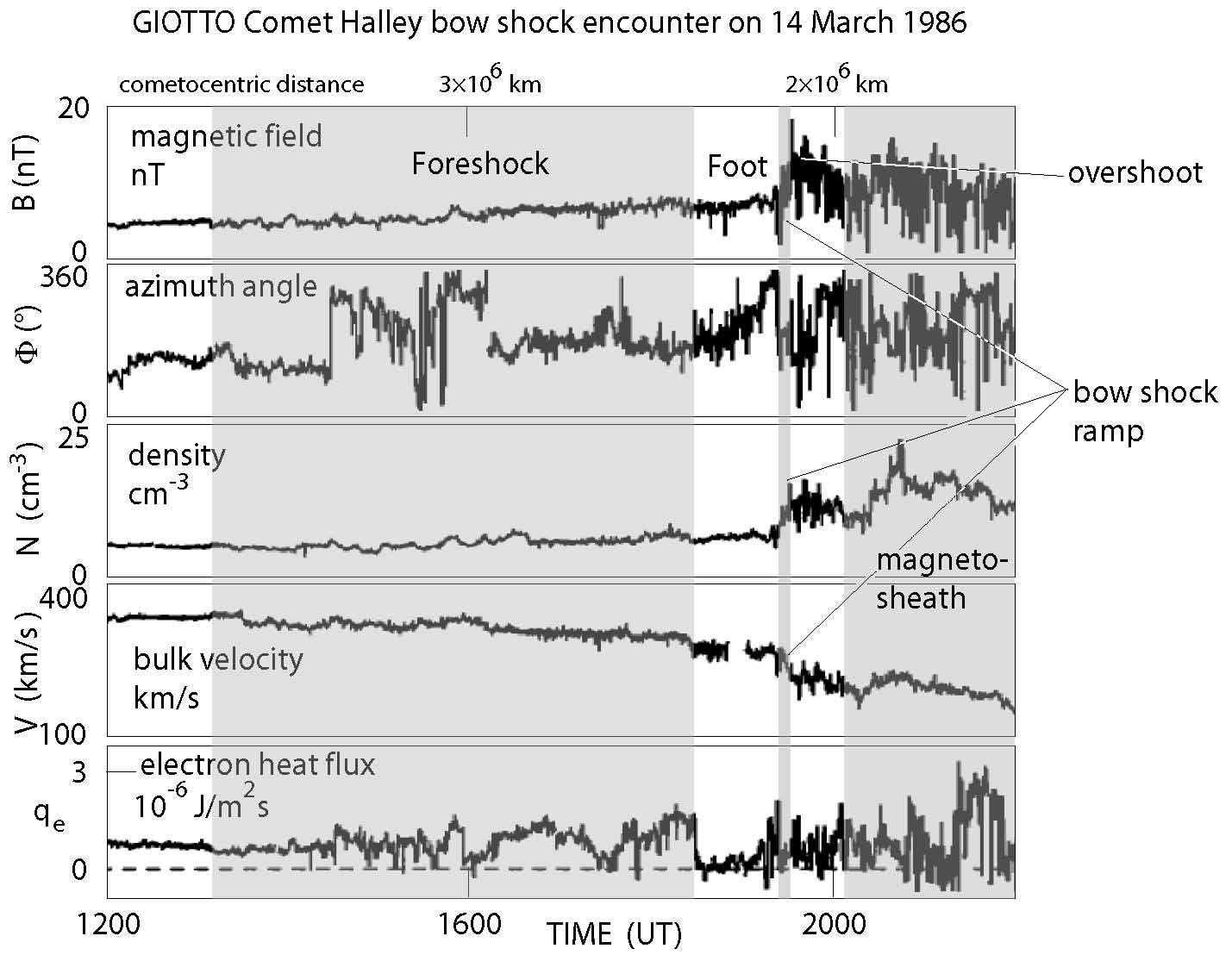}}
\caption[Giotto] 
{\footnotesize Giotto passage across the Halley bow shock on 14 March 1986 \citep[data taken from][]{Larson1992}. proving the existence of a cometary bow shock. The spacecraft entered the foreshock at about 1400 UT just before an interplanetary discontinuity passed over it which is seen in the turning of the magnetic field. After 1600 UT it was again in the foreshock, entering the foot region with the continuously incresing magnetic field and crossed the shock ramp at about $\sim$1930 UT. The shock compression factor was about 2, and the shock was oblique, neither parallel nor perpendicular. }\label{chapBS-fig-Giotto1}\vspace{-0.5cm}
\end{figure}

\subsubsection{Comet Halley}\index{shocks!Halley's bow shock}
\noindent The two most spectacular cometary events near the end of the past century were the destruction of comet Shoemaker-L\'evy when it hit Jupiter causing one of the greatest nuclear explosions in the heliosphere that could be observed from remote, and the human borne spacecraft visits to comet Halley. \index{spacecraft!Giotto}\index{comets!Halley}\index{comets!Shoemaker-L\'evy} 

Halley was visited by three spacecraft, the two Russian spacecraft {\small Vega 1}\index{spacecraft!Vega 1}\index{spacecraft!Vega 2} and {\small Vega 2}, and by {\small Giotto}. The {\small Vega} satellites reported the possible observation of a stand-off shock near Halley \citep{Galeev1986}. The existence of a bow shock at comet Halley was confirmed by the precise measurements of {\small Giotto} \citep{Neubauer1986}. This shock was crossed two times; on the inbound path, shown in Figure \ref{chapBS-fig-Giotto1}, the shock was found at cometocentric distance 1.2 Million km and was preceded by upstream waves. On the outbound crossing the shock was seen as a broad turbulent quasi-parallel shock. Figure \ref{chapBS-fig-Giotto1} shows a collection of the observations on 14 March 1986. At around $\sim$1930 UT {\small Giotto} crossed the shock ramp and entered the overshoot and magnetic pile-up magnetosheath region. It has before that time been in the foreshock where intense electron fluxes were detected flowing along the interplanetary magnetic field. Deceleration of the solar wind indicated entrance into the shock foot where the magnetic field started increasing and turning in direction while the electron heat flux decreased, an observation that coincides with what is known from planetary bow shocks. 

The shock ramp is seen in the steep increase in the magnetic field, strong deceleration of the flow, localised heat fluxes and violent changes in the direction of the magnetic field. The shock is oblique, probably more quasi-parallel than quasi-perpendicular. One of its peculiarities is that the transition takes quite a long time, much longer than on any of the magnetised planet shock crossing where the shock ramp is steep. Moreover, the determination of all these regions is not as clear as in the case of planetary bow shocks or even in the cases of Mars and Venus. The plasma and field quantities in the comet passage are more disturbed, take longer time to adjust to the downstream values, and the compression ratio is relatively small such that the shock is less strong than at a planet.  Nevertheless, the data show definitely that large comets like Halley do develop a bow shock in interaction between their gaseous environment and the solar wind when entering the inner heliosphere. However, the properties of the cometary shocks differ to a certain degree from the known planetary bow shocks. 
\begin{figure}[t!]
\vspace{-0.2cm}\centerline{\includegraphics[width=1.0\textwidth,clip=]{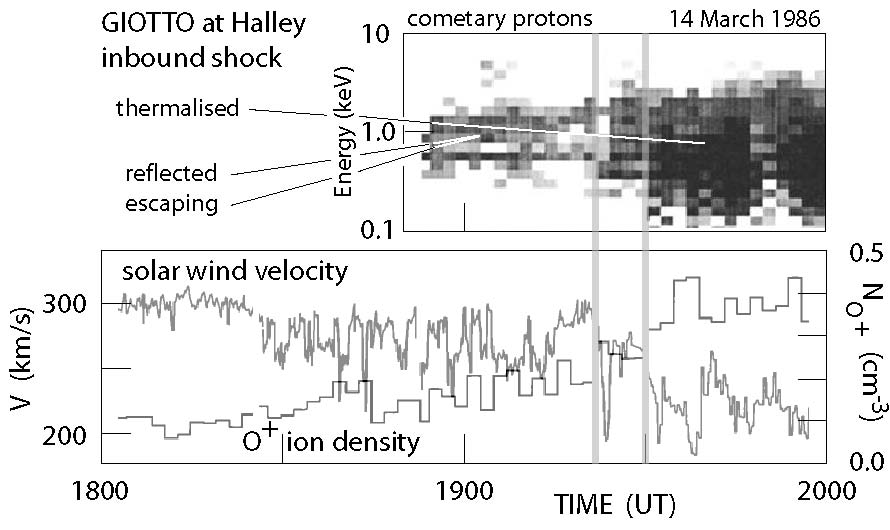}}
\caption[Giotto] 
{\footnotesize Giotto passage across the Halley bow shock on 14 March 1986 \citep[data taken from][]{Coates1990}. The lower panel shows the anticorrelation between the solar wind velocity and water ion content during approach of comet Halley and inbound shock crossing suggesting that mass loading by heavy cometary ions is responsible for the slowing down of the solar wind. The upper panel shows protons arriving at Giotto from the direction of Halley during approach of the comet. Cold reflected/escaping protons of energy $\sim$ 1 keV are detected prior to entering the foot and shock ramps, some of the ions have energies up to 5-6 keV, indicating acceleration either in the shock or foot region. Downstream the mean proton energy has decreased to $\sim$0.3 keV corresponding to the low mean bulk speed in the lower panel. At the smae time the proton distribution has been thermalised in the shock transition as see from the $\sim$1 keV spread of the distribution.}\label{chapBS-fig-Giotto2}\vspace{-0.5cm}
\end{figure}

\cite{Coates1990} have undertaken a thorough check of the various shock plasma parameters to confirm that the shock that was crossed by {\small Giotto} was indeed a shock. They found that the magnetosonic Mach numbers for the Halley shock were quite low, the inbound crossing during passage gave a Mach number of $1.1<{\cal M}_{ms}<1.8$, the Mach number of the outbound crossing was more stable $1.6<{\cal M}_{ms}<1.7$. Thus the Halley bow shock was quite weak and just marginally supercritical if the shock normal angle was small enough. The reason for this weakness may be found in the large number of heavy pick-up ions that have been loaded into the solar wind at and around shock location. Deceleration of the solar wind by this heavy mass loading clearly reduces the Mach number. This is shown in the lower panel of Figure \ref{chapBS-fig-Giotto2} where the solar wind velocity and the number density of cometary O$^+$-ions is plotted during shock approach and crossing. The anticorrelation of both quatities is obvious. The solar wind speed decreases gradually with increasing heavy cometary ion concentration during approaching the comet. The upper part of the figure shows the measured proton energy distribution for those protons which come from the direction of the comet. Before crossing the shock foot and ramp (shaded vertical bars) reflected or escaping protons are measured arriving from Halley. Note that these protons have low energy spread but relatively high energy $\sim$1 keV, part of them reaching energies of several keV indicating acceleration in the shock or shock foot region. (Heavy ion acceleration up to $<$250 keV at Halley has also been reported from downstream observations \citep{Kesc1992}. However, this acceleration is not necessarily due to the presence of the Halley bow shock.) After crossing the shock a broad thermal proton distribution of lower energy $<$1 keV is found that is typical for the downstream particle distribution.  \cite{Gombosi1991} determined the width $D$ of the region of heavy ion (O$^+$) outside the bow shock along the Giotto trajectory of being in the spatial interval between $2\times10^{\,6}\,{\rm km}>D>1.2\times10^{\,6}\,{\rm km}$. This thickness is determined by the ability of the particles to diffuse away from the comet.\index{shocks!mass loading}

Giacobini-Zinner and Halley have so far been the only comets from which {\it in situ} data have become available. Much theoretical and simulational work has been done to investigate the formation of shocks in front of a comet entering the inner heliosphere, work that was mostly based on magnetohydrodynamic considerations including cometary ions. Even the formation of a shock during Shoemaker-L\'evy 9 approach of Jupiter and its interaction with the Jovian bow shock has been simulated \citep{Lipatov1994} in order to show the deformation of the comet, ion heating and particle acceleration. Still, the formation of a shock around a comet has not yet been clarified satisfactorily. the observations suggest that sometimes, predominantly at larger comets, shocks develop. However their structure is different from planetary bow shocks because of mass loading, decrease in Mach number and the differences between the gaseous shell around a comet and the atmosphere of a planet like Venus.

\section{Conclusions}
\noindent This chapter has focussed on the discussion of the various planetary bow shocks that can form around the celestial bodies in the solar system and heliosphere. By the nature of the question this discussion was divided into a review of earth-like bow shocks, non-magnetised planetary bow shocks, atmospheric bow shocks, and finally cometary bow shocks. This division of the celestial objects in the solar system grouped the earth together with the tiny innermost planet, Mercury, and the giant outer planets, while the terrestrial planets Venus and Mars had to be attributed to the remaining groups. 

Interestingly enough, all planets and even the larger comets are surrounded by detached bow shocks. These bow shocks, for the magnetised planets, become ever stronger with increasing heliocentric distance from the sun, not only because of the increase in the internal magnetic fields from the inner planets Mercury and Earth to the giant outer planets but also because of the increasingly cooler and more dilute solar wind and the weaker magnetic field with distance from the sun. As a consequence the Mach number increases with distance, while the interplanetary magnetic field becomes more tangential to the planetary magnetic field. Thus the shocks become increasingly high Mach number quasi-perpendicular. In spite of this systematics, surprising differences have been found between the canonical paradigm of a planetary bow shock, Earth's bow shock, and the bow shocks of the outer giant planets. At Jupiter some of these differences could be attributed to the presence of the Jovian heavy plasmasphere and fast rotation leading to large centrifugal forces acting on the Jovian plasma and possibly affecting the shape and properties of Jupiter's bow shock. However, a general and interesting observation is that with increasing distance from the sun the bow shock obviously becomes narrower and steeper having a large overshoot and causing less plasma turbulence in the adjacent magnetosheath than expected. Also, the shock shape is more inflated the farther out from the sun we are. Finally, particle acceleration at the outer bow shocks seems to be more efficient than at Earth. This might be a result of the much broader magnetosheath at the giant planets which allows the turbulence to develop until quasi-stationarity.

Non-magnetised planets also possess bow shocks depending on whether they have atmospheres or not. The Moon's bow shock does not exist, however, as a global object since no induced or global lunar magnetic field has been found. However, locally, remanent magnetisation in the lunar crust sometimes leads to the formation of highly dynamic mini-bow shocks and magnetospheres. This is a very interesting finding that can extrapolated to the larger asteroids like Ceres but does, surprisingly, not apply to Mars. The reason is that Mars possesses a dilute atmosphere that is still dense enough to become ionised. Its ionosphere acts as a shield because the solar wind induces a magnetic pile-up of the magnetic field in front of it which then acts in a similar way as a planetary magnetic field even though being of purely interplanetary origin. Its existence is the reason for the formation of a detached Martian bow shock.

Venus with its very dense atmosphere, high surface temperature and proximity to the sun behave even differently. Venus is non-magnetic with even no remanent magnetism possible in its crust as its surface temperature is above the Curie point. However, its atmosphere becomes ionised, creates a dense ionosphere and leads to a pile-up of an induced magnetic barrier in a way similar to Mars. In front of this barrier and detached from it a bow shock stands close to Venus with similar properties as Earth's bow shock but the difference of being much smaller in size and being affected by atmospheric pick-up ions from Venus. 

Finally, comets can possess bow shocks as the example of comet Halley has shown. However, for this they need to enter the inner heliosphere for being sufficiently illuminated by the sun, outgas matter from their interior and generate a gas cloud that surrounds them. Such a cloud behaves similar to though not exactly like an atmosphere of a non-magnetic planet. It becomes ionised by illumination and charge exchange with the solar wind, creates an induced magnetic pile-up and thus a barrier which forces a bow shock to evolve if the comet is large enough. However, the mass loading of the solar wind substantially decreases the cometary Mach number such that the shock becomes weak and may even become subcritical not reflecting any particles. If this mass loading takes place over large distances from the comet, the solar wind might become gradually slowed down when approaching the comet and no shock will evolve at all even though a magnetic barrier forms around the comet. The solar wind then becomes sub-magnetosonic and has time enough to adjust for the presence of the comet and to flow around. This situation might have been observed at Giacobini-Zinner. In general, however, even large comets of the kind of Halley will develop a bow shock only inside a certain heliocentric distance from the sun. Farther out, even though the Mach numbers are large, the absence of a cometary atmosphere prevents its formation. The only possibility for a shock then is that the fast solar wind stream induces an inductive field in the cometary body and conducting cover layer. This however requires sufficiently large cometary nuclei of size substantially exceeding the local proton gyroradius in the solar wind stream.

{\acknowledgements  This work has been part of an ISSI Working Group effort on ``The Physics of Collisionless Shocks in the Heliosphere". C. H. Jaroschek was supported by grants from the Japanese Society for the Promotion of Science which he gratefully acknowledges. We thank Masahiro Hoshino (Tokyo University)  for his continuous support and helpful discussions. R.T. thanks Andr\'e Balogh (ISSI) for his moral support and continuous interest.}

\end{document}

%% file: calbook_a_defs.tex
\def\3{\ss}

\def\q0{\phantom{1}}

\def\thetabn{\Theta_{Bn}}

\def\parp2{\frac{\partial^{2}}{\partial p^{2} }}

\def\m21{$2^{\circ}\times 1^{\circ}$}

\def\ts{\thinspace}

\def\ne8{Ne\ts{$\scriptstyle {\rm VIII}$} }

\newcommand{\aap}{{Astron.\ Astrophys. }}

\newcommand{\asr}{{Adv. \ Space. \ Res.}}

\newcommand{\grl}{{Geo\-phys.\ Res.\ Lett.}}

\newcommand{\jgr}{{J.\ Geo\-phys.\ Res.}}

\newcommand{\pss}{{Planet.\ Space Sci.}}

\newcommand{\ssr}{{Space Sci.\,Rev.}}

\def\ion[#1 #2]{#1\,{\sc #2}}
\def\lamb[#1]{#1\,{\AA}}
\def\serts89{SERTS-89}
\def\fe12{Fe\,{\sc xii}}

\def\mb[#1]{\makebox[0.15cm][l]{#1}}
